\documentclass[traditabstract]{aa} 
\usepackage{graphicx}
\usepackage{natbib}
\usepackage{txfonts}
\newcommand{\ergs}{\>{\rm erg}\,{\rm s}^{-1}}
\newcommand{\kms}{$\rm{\,km \,s}^{-1}$}
\newcommand{\ergscmA}{\>{\rm erg}\,{\rm s}^{-1}\,{\rm cm}^{-2}\,{\rm \AA}^{-1}}
\begin{document}
   \title{Looking for the least luminous BL Lac objects.}

   \subtitle{}

   \author{A. Capetti\inst{1}
          \and
          C. M. Raiteri\inst{1}
          }

   \institute{INAF-Osservatorio Astrofisico di Torino, via Osservatorio 20, I-10025 Pino Torinese, Italy}

   \date{}

  \abstract {Among active galactic nuclei, BL Lac objects show extreme
    properties that have been interpreted as the effect of relativistic
    beaming on the emission from a plasma jet oriented close to the line of
    sight. The Doppler amplification of the jet emission makes them 
    ideal targets for studying jet physics. In particular, low-power BL Lacs
    (LPBL) are very interesting because they probe the jet formation and
    emission processes at the lowest levels of accretion.  However, they are
    difficult to identify since their emission is swamped by the radiation
    from the host galaxy in most observing bands.

In this paper we propose a new LPBL selection method based on the mid-infrared
emission, in addition to the traditional optical indices. We considered the
radio-selected sample of Best \& Heckman (2012, MNRAS, 421, 1569) and
cross-matched it with the {\it WISE} all-sky survey. In a new diagnostic plane
including the W2$-$W3 color and the Dn(4000) index, LPBL are located in a
region scarcely populated by other sources. By filtering objects with
small emission line equivalent width, we isolated 36 LPBL candidates up to
redshift 0.15. Their radio luminosity at 1.4 GHz spans the range $\log L_{\rm
  r} = 39.2$--41.5 [$\rm erg \, s^{-1}$].

Considering the completeness of our sample, we analyzed the BL Lac
luminosity function (RLF), finding a dramatic paucity of LPBL with respect to
the extrapolation of the RLF toward low power. This requires a break in the
RLF located at $\log L_{\rm r} \sim 40.6$ [$\rm erg \, s^{-1}$]. The
consequent peak in the BL Lacs number density is possibly the manifestation of
a minimum power required to launch a relativistic jet.}

   \maketitle

\section{Introduction}

BL Lac objects (BL Lacs), together with flat-spectrum radio quasars (FSRQs),
form a class of active galactic nuclei (AGNs) known as ``blazars".  Like the
other radio-loud AGNs, they are thought to be powered by a central
supermassive black hole, surrounded by an accretion disk.  Two plasma jets are
ejected in the direction orthogonal to the disk.  Among radio-loud AGNs,
blazars show extreme properties, such as strong flux variability on the whole
electromagnetic spectrum even on intra-day time scales \citep{wag96}, variable
optical and radio polarization \citep{all96,smi96}, and apparent superluminal
motion of radio components \citep{kel04}. The explanation of this observing
evidence requires that one of the relativistic plasma jets be closely aligned
with the line of sight, causing Doppler beaming of the jet emission
\citep{urr95}. This makes blazars the ideal candidates for studying the physics of
AGN jets. In particular, the least luminous blazars are extremely interesting
because they can probe the jet formation and emission processes at the lowest
levels of accretion.

The classical distinction between BL Lacs and FSRQs is based on the strength
of the emission lines in the optical spectrum, with BL Lacs being identified
by rest-frame equivalent widths (EW) less than 5 \AA\ \citep{sti91}.  The
blazar spectral energy distribution (SED) is dominated by the non-thermal
emission from the jet. It shows two bumps: the low-energy one (from radio to
UV--X-rays) is due to synchrotron radiation, while the high-energy bump (X-
and $\gamma$-rays) is probably due to an inverse-Compton process off the same
relativistic electrons that produce the synchrotron photons \citep{kon81}.
Depending on the frequency of the synchrotron peak $\nu_{\rm p}$ in the SED,
BL Lacs have been divided into low-, intermediate-, and high-energy
synchrotron peaked BL Lacs. The objects of the first class (named LBL or LSP)
have $\nu_{\rm p} < 10^{14} \rm \, Hz$. For those of the second class (IBL or
ISP) $10^{14} < \nu_{\rm p} < 10^{15} \rm \, Hz$, and for objects of the third
class (HBL or HSP), $\nu_{\rm p} > 10^{15} \rm \, Hz$ \citep{abdo2010}. It has
been claimed that blazar SEDs follow a ``blazar sequence"
\citep{fos98,donato01}, linking the SEDs of FSRQs at the highest synchrotron
peak luminosities $L_{\rm p}$ and lowest $\nu_{\rm p}$ to the SEDs of LBL,
IBL, and finally HBL, when progressively decreasing $L_{\rm p}$ while
increasing $\nu_{\rm p}$.  This view has been criticized, for example, by
\citet{gio12a}, who underline the selection effects of flux-limited
surveys and lack of redshift determination for the likely high-$L_{\rm
  p}$-high-$\nu_{\rm p}$ sources.

The difficulty of finding low-luminosity BL Lacs comes from the dilution of
the jet component by the host-galaxy emission in the optical and near-IR
bands, as well as from the flux limit of the available surveys. As a result,
their radio luminosity function (RLF) is only determined above $\log L_{\rm r}
\sim 40.5$ [$\rm erg \, s^{-1}$] \citep[e.g.,][]{wol94,pad07}.  The analysis of
\citet{marcha2013} reaches lower luminosities when also including ``Type 0''
and weak-lined radio galaxies. Our aim is to find a substantial number of LPBL
and to build their RLF down to a poorly studied regime of radio power.

   \begin{figure*}
    \includegraphics[width=95mm]{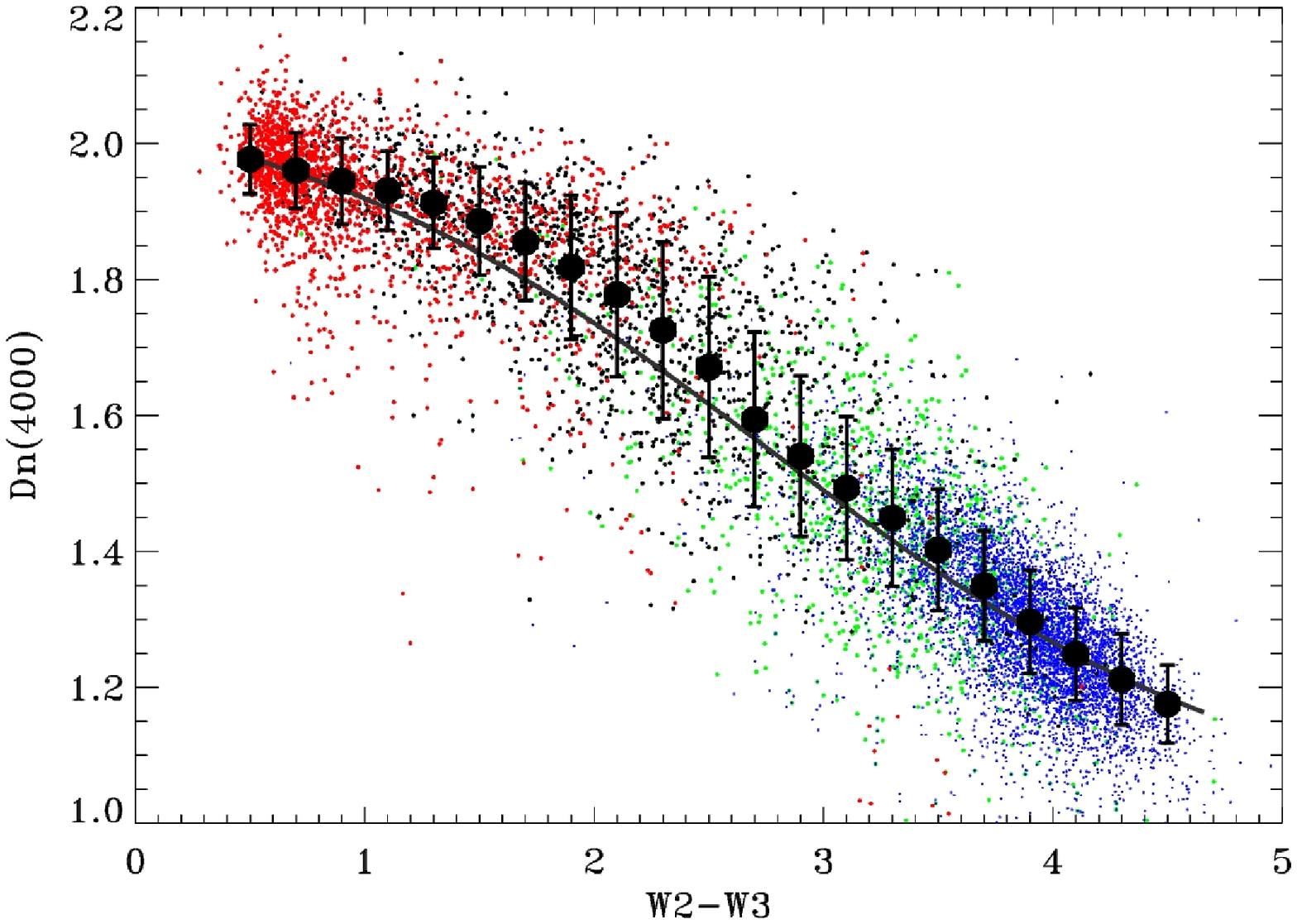}
    \includegraphics[width=95mm]{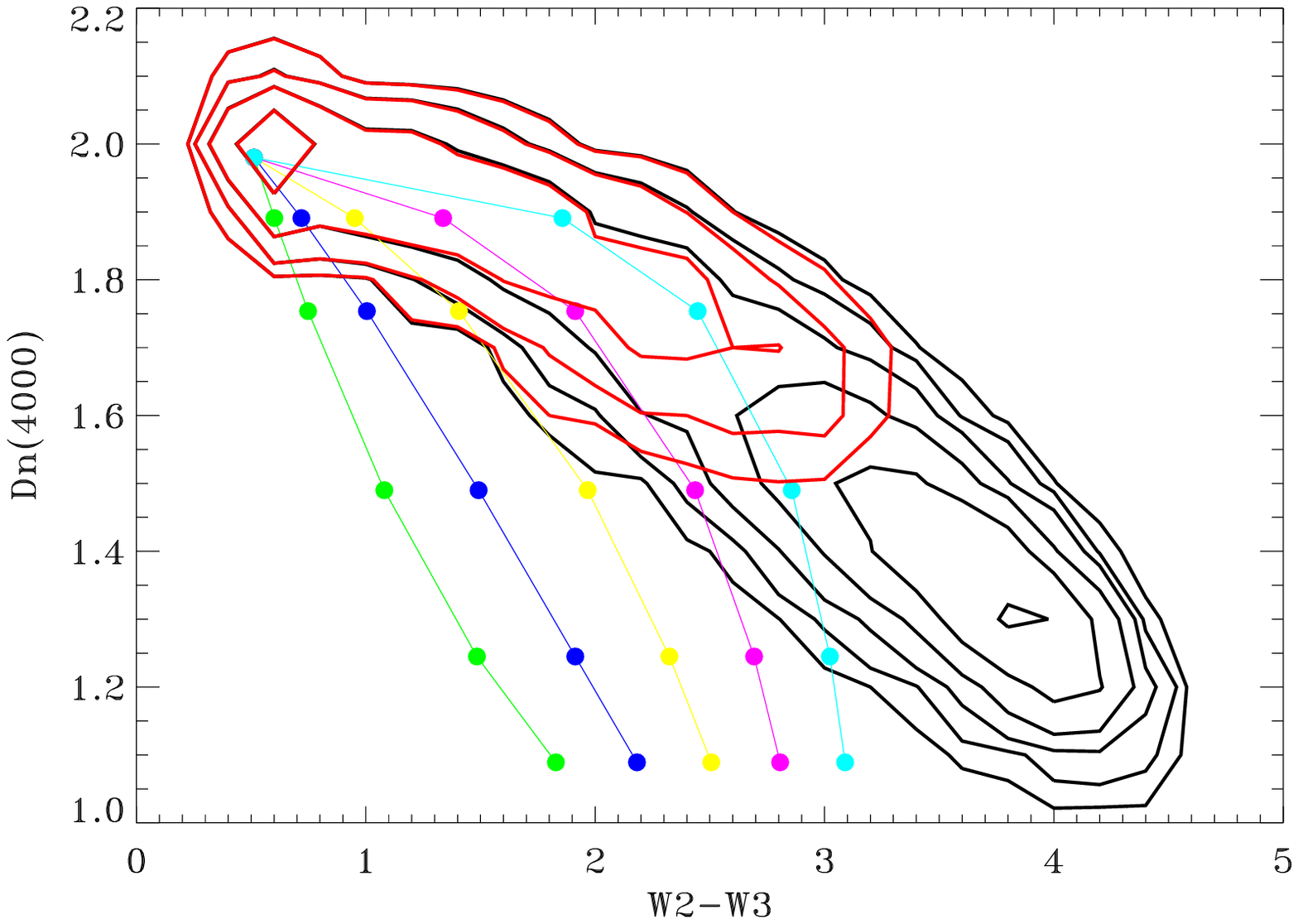}
    \caption{Left. Relation between the Dn(4000) index and the W2$-$W3 color
      measured from the sample of 21~065 bright nearby galaxies (see text for
      the selection criteria). For each bin in W2$-$W3, the large black dots
      denote the mean value and dispersion of the best Gaussian fit to the
      Dn(4000) distributions. Each spectroscopic class is highlighted with a
      color: red for lineless galaxies, black for LINERs, green for Seyferts,
      and blue for star-forming galaxies.  The solid line represents the
      result of a toy model where a fixed fraction of a blue light component
      of increasing strength is reprocessed into MIR radiation.
      Right. Simulated Dn(4000) versus W2$-$W3 tracks obtained by adding a
      power-law component of increasing strength with a spectral index
      increasing from 0 (green) to 1 (cyan) with a step of 0.25 to the SED of
      a quiescent galaxy. Within each track we label the ratios of 0, 0.1,
      0.3, 1, 3, and 10 between the jet and galactic components. The black
      lines trace the iso-density contours of the distribution shown in the
      left panel; the red lines show the iso-densities contours for the
      objects with all lines rest frame EW $<$ 5 \AA.}
   \label{uvmir}
   \end{figure*}

This paper is organized as follows. In Sect.\ 2 we present a new selection
method for identifying BL Lacs that involves both optical and MIR
properties. This leads to singling out 36 BL Lac candidates with redshift $z \le
0.15$ and radio luminosities in the range $\log L_{\rm r} \sim 39$ to 42
[$\ergs$] in Sect.\ 3. The various sources of incompleteness in our sample are
considered in Sects.\ 4 and 5, while the resulting BL Lac RLF is derived and
also discussed in the framework of the AGN unification scheme in
Sect.\ 6. After a summary, our conclusions are drawn in Sect.\ 7.

Throughout the paper we adopt a cosmology with $H_0=70 \, \rm km \, s^{-1} \,
Mpc^{-1}$, $\Omega_{\rm M}=0.29$, and $\Omega_\Lambda=0.71$.  When we speak of
radio luminosities $L_{\rm r}$, we mean $\nu_{\rm r} l_{\rm r}$, where the
monochromatic luminosity $l_{\rm r}$ is measured at $\nu_{\rm r}=1.4$
GHz. Moreover, we distinguish between radio luminosities in cgs units, $L_{\rm
  r}$, and those in MKS units, $P_{\rm r}$.

\section{The selection method}
\label{selection}

   \begin{figure*}
   \includegraphics[width=95mm]{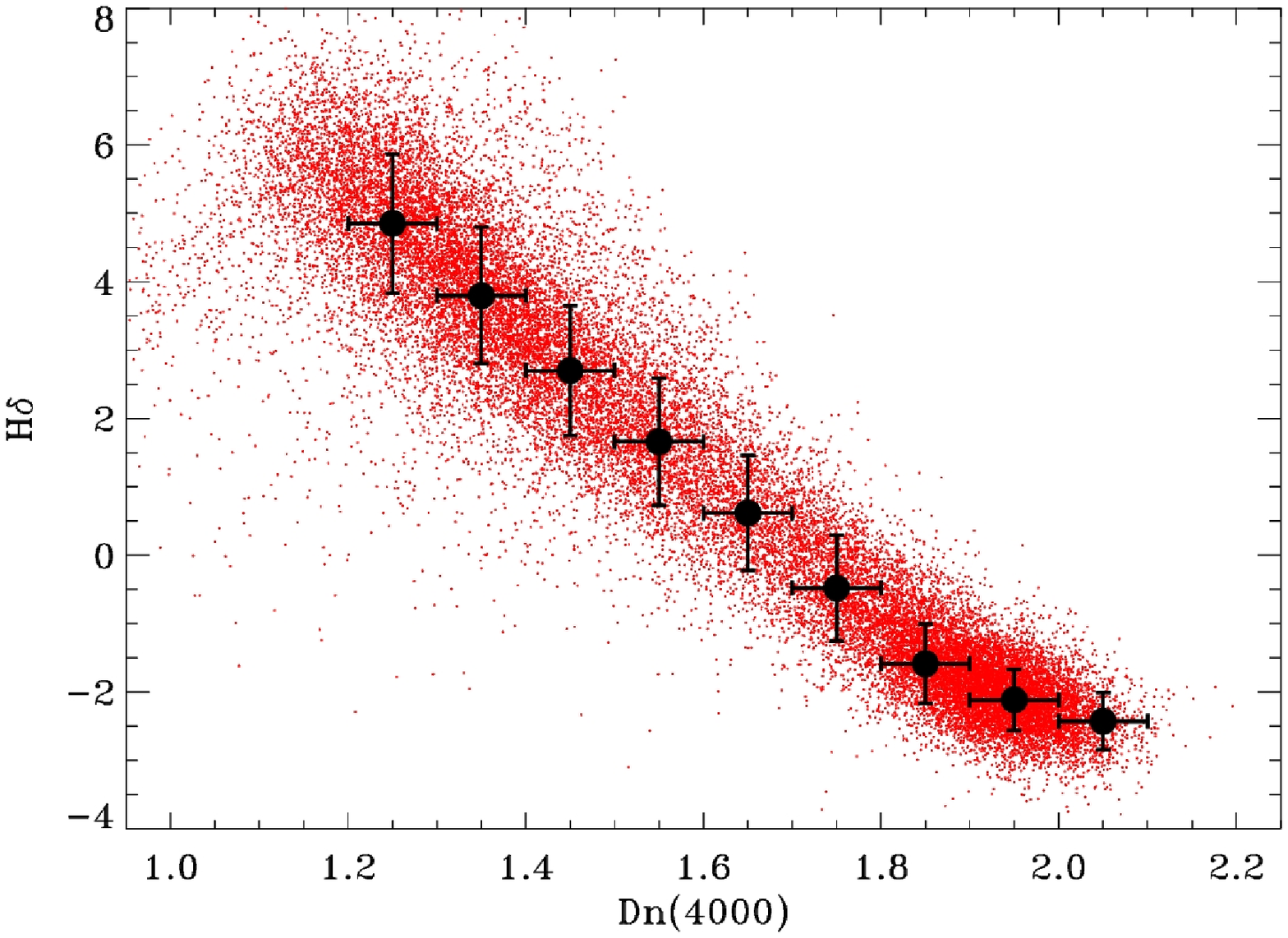}    
   \includegraphics[width=95mm]{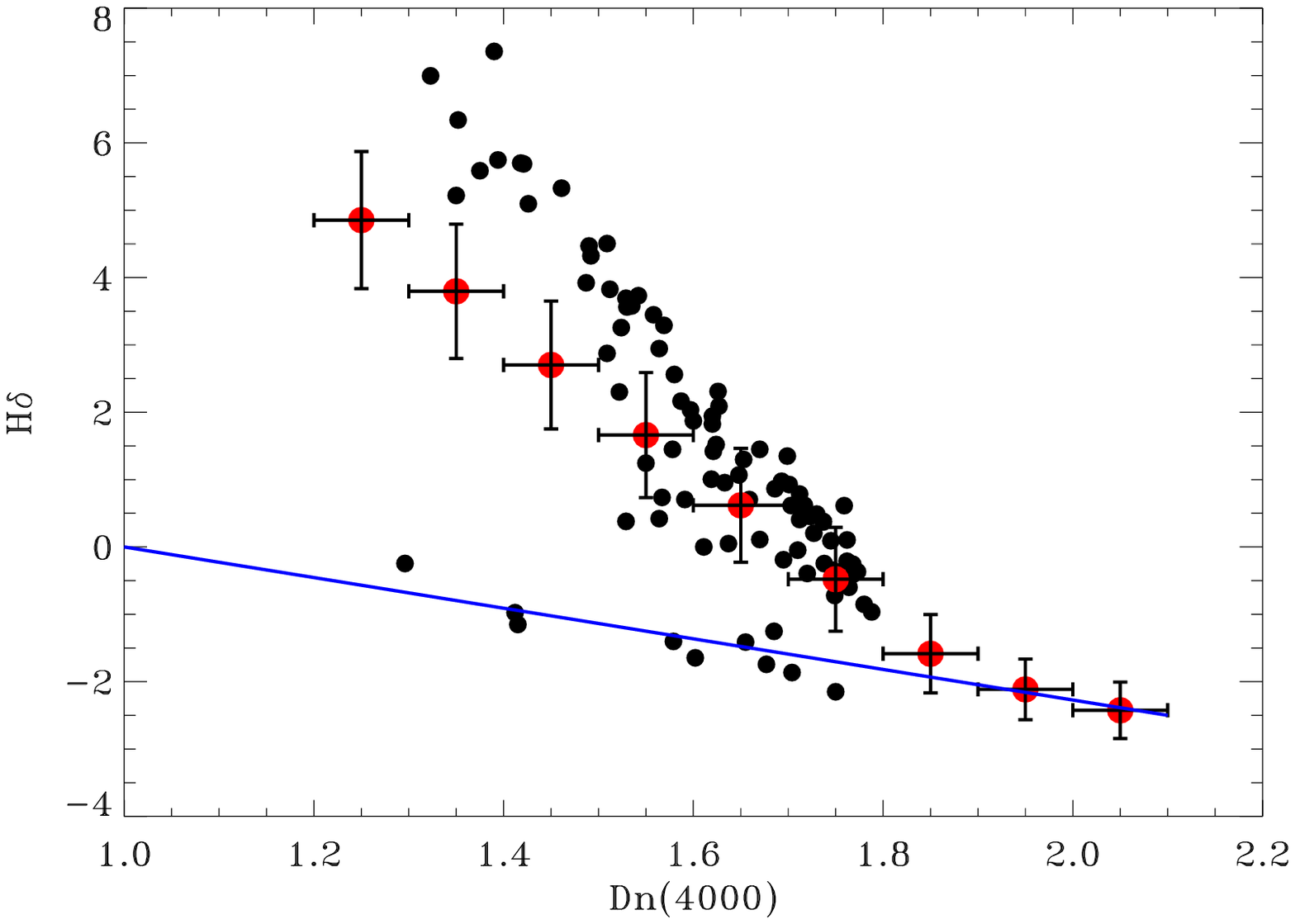}    
   \caption{Left. Relation between the Dn(4000) and H$\delta$ indices from the
     SDSS subsample of 21~065 galaxies. For each bin in Dn(4000), the black
     dots indicate the mean value and dispersion of the best Gaussian fit to the
     H$\delta$ distribution.  Right. H$\delta$ versus Dn(4000) for the
     outliers of the Dn(4000) versus W2$-$W3 relation (black dots). The red
     dots label, for each bin in Dn(4000), the mean value and dispersion of the
     best Gaussian fit to the H$\delta$ distribution for the overall SDSS
     galaxies population. The straight blue line is the expected locus of BL
     Lacs, obtained by varying the ratio between jet and galactic emission
     (see text for details). }
    \label{hddn}
   \end{figure*}

As mentioned, the peculiarity of BL Lacs is their non-thermal emission from a
relativistic jet seen at a small angle with respect to the line of sight.  As
a consequence, their optical spectrum is either featureless or only presents
emission lines of small EW. This is because lines (when present) are
isotropic, while the jet emission is highly beamed.  Similarly, the jet
dilutes the stellar absorption features.  BL Lacs have been historically
selected by setting a limit of 5 \AA\ to the rest-frame EW of any emission
line \citep{sti91}.  Alternatively, they have been recognized as objects with
low Dn(4000) index\footnote{Defined according to \citet{bal99} as the ratio
  between the flux density measured on the ``red" side of the Ca~II break
  (4000--4100 \AA) and that on the ``blue" side (3850--3950 \AA).}.  The
Dn(4000) limit was set to 1.33 by \citet{sto91}, then later relaxed to 1.67 by
\citet{marcha1996} and \citet{plo10}.\footnote{ Various authors use different
  definitions of the Ca II break strength. Our parameterization is related to
  those found in the literature as ${\rm Dn}(4000) = 1 / (1 - C)$
  \citep{landt02}.}  Nevertheless, passive elliptical galaxies have
Dn(4000)$\sim$2 (see below), so that we can expect that the Dn(4000)
constraint can be further relaxed to find LPBL \citep[see also][]{marcha2013}.

We propose to loosen the Dn(4000) limit but to include a further selection
tool based on the mid-infrared emission (MIR).  In fact, in active galaxies
the MIR is also an isotropic quantity, owing to the reprocessing of (mainly)
UV photons from the interstellar dust, while in BL Lacs it also receives a
non-thermal contribution from the jet that alters the overall SED shape.  As
we discuss in detail below, the resulting MIR properties enable us to
separate BL Lacs from other classes of active galaxies.

The {\it WISE} satellite provides us with an all-sky survey in the four bands W1,
W2, W3, and W4, which are centered at 3.4, 4.6, 12, and 22 $\mu$m,
respectively \citep{wri10}. The best-suited color for our purposes is obtained
from the W2 and W3 bands, because it has high wavelength leverage and level
of completeness.

Before we can proceed to use the MIR observations for the BL Lacs selection, we
must explore the relationship between the optical and MIR properties of galaxies in general.
For this purpose, we consider the 927~552 SDSS DR 7 spectra of the 818~333 galaxies (MPA-JHU sample hereafter) in the value-added spectroscopic catalogs produced by the group from the Max Planck Institute for Astrophysics, and The Johns Hopkins University, and available at {\tt http://www.mpa-garching.mpg.de/SDSS/} \citep{bri04,tre04}.

Our aim is to detect the faintest BL Lacs, so we started to analyze galaxies
up to $z=0.1$ and considered only massive objects, since they are the typical
BL Lac hosts.  We then filtered objects with high-quality data
with the following criteria:
\begin{itemize}
\item[i)] redshift $z \le 0.1$, 
\item[ii)] error on Dn(4000) $<$ 0.02, 
\item[iii)] signal-to-noise (S/N) in both W2 and W3 bands $>10$, 
\item[iv)] stellar velocity dispersion $>$ 100 \kms.
\end{itemize}

In Fig. \ref{uvmir} (left panel) we compare the W2$-$W3 color with the
Dn(4000) index for the 21~065 selected galaxies.  The presence of a clear
sequence emerges, with the Dn(4000) value decreasing with increasing W2$-$W3.
In each W2$-$W3 bin, we estimated the mean value and dispersion of the best
Gaussian fit to the Dn(4000) distribution.

Taking advantage of the SDSS spectra, we can consider the various
classes of galaxies separately. They are located in different areas of this diagram:
the passive galaxies are clustered at the top left, with Dn(4000) $\sim 2$
and W2$-$W3 $\sim$ 0.5. Proceeding toward the opposite corner of the diagram we
find LINERs, most of them having Dn(4000) $\sim$ 1.5--2.0 and 1 $<$ W2$-$W3 $<
3$, and then Seyfert galaxies with Dn(4000) $\sim$ 1.5 and 3 $<$ W2$-$W3 $<$
4. In the bottom right hand corner, we find star-forming galaxies, with low
Dn(4000) values ($\sim 1.2$--1.3) and large excesses in the W3 band. Therefore,
all spectroscopic classes of galaxies follow the same overall sequence possibly
because, regardless of the origin of the blue/UV light, the same mechanism of
reprocessing into the MIR is operating.

To support this interpretation further we built a very simple model. We added
a growing contribution
of blue light to the SED of a passive galaxy and adopted a constant reprocessing
efficiency into MIR emission. The track obtained is shown in Fig.\ \ref{uvmir} (left panel) and follows the central values of the distributions remarkably well.

It is important to note that the same sequence is followed by the
  objects with low values of the emission lines EW, the sources among which we
  expect, in principle, to find BL~Lacs. By setting a limit to the lines rest
  frame of EW $<$ 5 \AA, we obtain the red contours overplotted in
  Fig.\ \ref{uvmir}. The distribution of this subsample follows that
  of the general galaxies population closely up to W2$-$W3 $\sim$ 3. The low EW objects
  only avoid the region of largest mid-infrared colors, which are typical of star-forming
  galaxies.

We can now explore the expected behavior of BL Lac objects in the same
diagram.  This was simulated by adding to the SED of a quiescent galaxy a jet
component $F_{\rm jet}$.  For the quiescent galaxy we use a Dn(4000)=1.98 and
W2$-$W3=0.5, i.e.\ the coordinates of the upper left average point in
Fig.\ \ref{uvmir}, while for the jet component we adopt a power-law spectral
shape in the MIR--optical frequency range of the form $F_{\rm jet} = F_0 \,
\nu^{-\alpha}$.  When $F_{\rm jet}$ increases, the W2$-$W3 color grows, owing to
the larger relative contribution of $F_{\rm jet}$ in the W3 band than in the
W2 band with respect to the galactic emission. At the same time, an excess of
blue light emerges, thus reducing the Dn(4000) index (see Fig.\ \ref{uvmir},
right panel).  Depending on the value of $\alpha$, the expected locus of BL
Lacs varies. The curves obtained with different values of $\alpha$ can be
overplotted onto the relation between W2$-$W3 and Dn(4000) obtained above.
All the simulated tracks are steeper than the phenomenological trend traced by
the sample considered above. As a result, the expected BL Lac tracks exit the
region where the SDSS galaxies are located. The transition into the
  ``forbidden zone'' occurs at Dn(4000) $\sim$ 1.8 for the flatter BL Lac SED
  and at lower values for softer SEDs. The visual inspection of broad-band
  SEDs of various classes of BL~Lacs (e.g., \citealt{giommi12,rai14})
  indicates that an optical-to-MIR index $\alpha\sim0.5$ is typical of HBL,
  while $\alpha\sim1$ can be associated to LBL.

In other words, the presence of a BL Lac nucleus produces a
substantial excess of blue light that corresponds to a smaller MIR excess than
in the case of active or star-forming galaxies.
We conclude that this diagnostic plane can be used to isolate BL Lac objects.

\subsection{Outliers in the Dn(4000) versus W2$-$W3 plane}
\label{outliers}

\begin{figure*}
\sidecaption
\includegraphics[width=12cm]{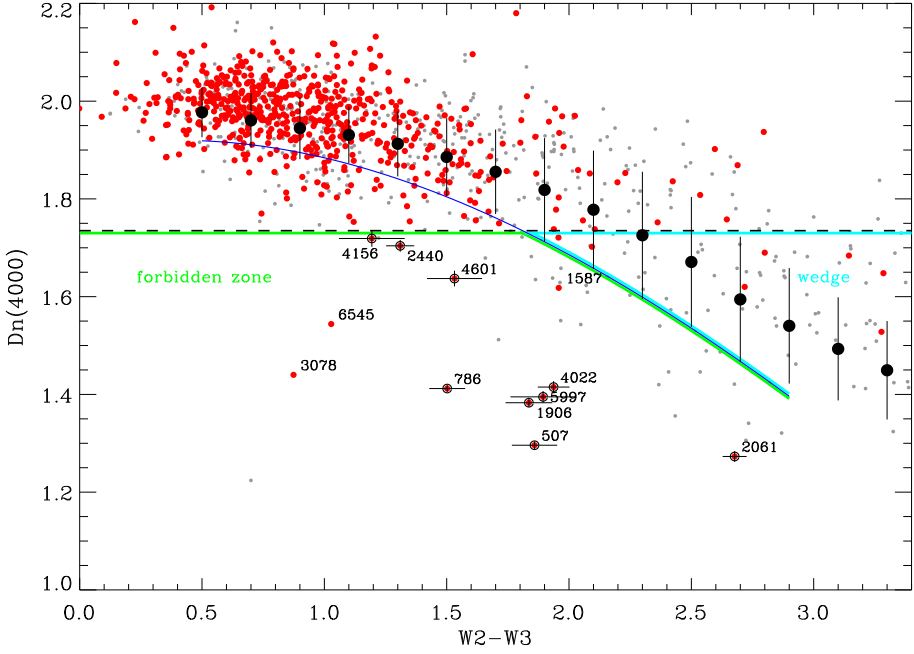}
\caption{Dn(4000) versus W2$-$W3 diagram. Black filled circles represent the
  relationship derived for the MPA-JHU sample that is shown in
  Fig.\ \ref{uvmir}. The blue line is a polynomial fit to the 1-$\sigma$ limit
  of this relationship. Gray dots represent the sources in the BH12-A sample
  with $z \le 0.1$ and S/N$>2$ in the W2 and W3 bands. Red dots highlight the
  objects with rest-frame emission line EW$<5$ \AA\ and a FIRST central
  counterpart.  Red dots circled in black are the nine BL Lacs candidates,
  i.e.\ those sources with $\rm Dn(4000) < 1.735$ (shown by the horizontal
  dashed line) that meet the additional constraint on the $H\delta$ index (see
  Fig.\ \ref{best_lick_z1}). The boundaries of the ``forbidden" zone and `wedge" 
  region are highlighted with a green and a cyan line, respectively.}
\label{dn_wise_z1}
\end{figure*}

Although SDSS galaxies define a clear locus in the Dn(4000) versus W2$-$W3
plane, several outliers are present in the region where we expect to find BL
Lacs. These objects thus deserve further investigation in order to ensure the
purity of the sample.  We then individually consider the 112 objects
(representing only 0.5\% of the 21~065 selected galaxies) lying below the
average Dn(4000) versus W2$-$W3 relation by more than 3$\sigma$. In nine cases
a bright star is located close to the galaxy's center and is covered by the
SDSS aperture; this causes the reduced Dn(4000) value. In 11 objects we are
dealing with clear misidentifications, since they are all high-redshift QSOs
($z=1.5$--2.3).

There are still 92 objects. When looking at their SDSS spectra, it emerges that
most of them have a pronounced absorption corresponding to the Balmer
lines. These objects are often referred to as E+A galaxies \citep{got03}. 
The presence of the A stars produces a substantial contribution in the 4000 \AA\ spectral
region, leading to low Dn(4000) values. The lack of the corresponding increase
in the {\it WISE} color is due to the spectral shape of the A stars, with a sharp
drop below the Balmer break. The paucity of UV photons, which are most
effectively absorbed by dust and are consequently the dominant mechanism of
dust heating, accounts for the reduced MIR emission. Nonetheless, such sources
can be easily recognized by measuring their H$\delta$ index \citep{worthey97}.

In Fig. \ref{hddn} (left panel) we show the relation between the Dn(4000) and H$\delta$
indices for the same sample of SDSS galaxies as used above. The H$\delta$
index monotonously decreases with increasing Dn(4000).

In Fig.\ \ref{hddn} (right panel), we plot the 92
objects located in the ``forbidden zone" in
Fig. \ref{uvmir} in the same diagram. They clearly separate into two sequences. The first follow
the general H$\delta$ versus Dn(4000) trend, and they are most likely E+A galaxies.
Their position within the sequence depends on the fractional contribution of A stars in each given object. The second group of objects is instead located where BL
Lacs are expected. This region connects the bottom right hand corner
of the diagram, which is characteristic of quiescent galaxies, to the location of a
featureless source, having Dn(4000) = 1 and H$\delta$=0. Indeed, five out of ten
of them are associated with a radio source with a flux density higher than $\sim 50$
mJy, which is a clear indication for the presence of non-thermal emission. 
We recover them as BL Lac candidates in Sect.\ \ref{sase}.

Summarizing, the proposed tool for isolating BL Lacs from a large population of
galaxies based on the Dn(4000) versus {\it WISE} W2$-$W3 color diagram appears to be
promising, particularly for HBL objects, which are expected to populate a
``forbidden zone" in this diagram.  However, the presence of outliers in this
region requires 
\begin{itemize}
\item[i)] visually exploring the SDSS data to remove
objects whose spectrum is contaminated by a nearby star,
\item[ii)] individually checking for misidentified objects,
\item[iii)] most importantly, including a further diagram formed by the Dn(4000) and H$\delta$ indices in
the analysis in
order to exclude E+A galaxies. 
\end{itemize}
After such an analysis we are left with only five
possible contaminants to the BL Lacs population, i.e.\ five galaxies (out of 21~065) with a low value of Dn(4000) but lack both MIR excess and radio emission. 
The nature of these objects is unclear; however, due to their extreme rarity, we do not expect them to significantly affect the results of our study.

\section{Looking for the least luminous BL Lacs}
\label{sase}

\begin{figure}
\resizebox{\hsize}{!}{\includegraphics{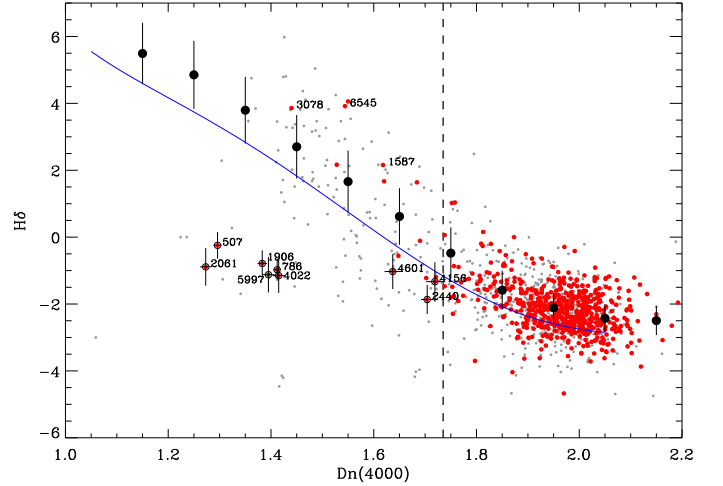}}  
\caption{$\rm H\delta$ index versus Dn(4000) diagram. Black filled circles represent the relationship derived for the MPA-JHU sample that is shown in Fig.\ \ref{hddn}. The blue line is a polynomial fit to the 1-$\sigma$ limit of this relationship. All objects above this line are considered E+A contaminants. Gray dots refer to the sources in the BH12-A sample with $z \le 0.1$ and S/N$>2$ in the W2 and W3 bands. Red dots highlight the objects with the emission line EW$<5$ \AA\ and a FIRST central counterpart. Red dots circled in black are the nine BL Lacs candidates. The vertical dashed line indicates the limit $\rm Dn_{max}=1.735$.}
\label{best_lick_z1}
\end{figure}

We looked for the least luminous BL Lacs by considering the sample of 18~286
radio sources built by \citet[][hereafter BH12]{bes12alias}. They combined
the MPA-JHU sample with the NVSS and the FIRST
surveys, the cross-matching going down to a flux density level of 5 mJy in the
NVSS.  We first focused on the 7302 objects they classified as radio-AGNs,
separating them from the star-forming galaxies.  These authors
also restricted their work to radio sources within the ``main galaxy sample"
\citep{str02}, comprising those galaxies with magnitudes in the range $14.5 <
r < 17.77$ and redshift range $0.01 < z < 0.3$.
The 7302 sources (hereafter the BH12-A sample) were cross-matched with the AllWISE catalog available at {\tt http://irsa.ipac.caltech.edu/} using a 3 arcsec search radius and finding 7242 counterparts\footnote{Actually 7271, but 29 are double identifications of which we kept only the closest object}.

Since our aim is to detect the faintest BL Lacs, we started to analyze the closest Universe, up to $z=0.1$.
In Fig.\ \ref{dn_wise_z1} we present the Dn(4000) versus W2$-$W3 plot for all objects with $z \le 0.1$ and with S/N of {\it WISE} data in the W2 and W3 bands $> 2$. A comparison with Fig.\ \ref{uvmir} shows that most radio galaxies are located in the same area as the MPA-JHU sample and that we are now dealing with mostly weakly active galaxies, with $\rm Dn(4000) \sim 2$ and W2$-$W3 in the 0.5--1.5 range, while the locus of star-forming galaxies is poorly populated.
When further selecting objects with i) rest-frame EW of the main emission lines $< 5 \AA$ and ii) existence of a FIRST central counterpart to exclude double sources lacking a radio core, we are left with 609 sources (Fig.\ \ref{dn_wise_z1}), only a few of which lying in the ``forbidden zone".

We searched for BL Lac candidates among those selected objects whose ratio
between the jet to galactic emission at 3900 \AA\ is $f > 1/3$, corresponding
to ${\rm Dn=(Dn^0}+f)/(1+f)<1.735$, having adopted a Dn(4000) index for
quiescent galaxies $\rm Dn^0=1.98$ (see the previous section)\footnote{
    Dn(4000)=1.735 corresponds to $C=0.423$.}.  The choice of the $f$ lower
limit is arbitrary, since it is a compromise between the possibility of detecting a
non-thermal signature (hence distinguishing a BL Lac candidate from a
passive galaxy) and the requirement of completeness. However, this assumption
will be included self-consistently in the analysis presented in the following
sections.
In Fig.\ \ref{dn_wise_z1} there are 12 objects with Dn(4000) $< \rm Dn_{\rm max}= 1.735$ and at the same time below the 1-$\sigma$ limit traced by the MPA-JHU sample (see Sect.\ \ref{selection}). These are labeled with an ID number.

Figure \ref{best_lick_z1} shows the $\rm H\delta$ versus Dn(4000) diagram. 
If we require that $\rm H\delta$ is lower than the 1-$\sigma$ limit traced by the MPA-JHU sample to exclude contaminating E+A galaxies (post-starburst) as explained in Sect.\ \ref{selection}, we are left with nine sources. Their properties are listed in Table \ref{bzall}.
Six of them are known\footnote{They are included in the catalog of blazars ``Roma-BZCAT" \citep{mas09}, available at {\tt http://www.asdc.asi.it/bzcat/}.} BL Lacs with $\rm Dn(4000) \sim 1.3$--1.4, while the other three are new candidates with Dn(4000) values of about 1.7. 

\begin{figure}
\resizebox{\hsize}{!}{\includegraphics{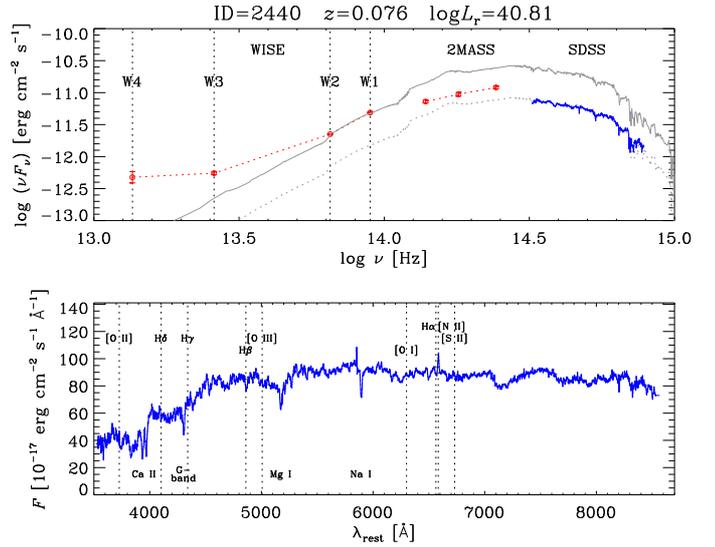}}  
\caption{Top: SED of a new BL Lac candidate, ID=2440. Red circles display {\it WISE}
  and dereddened 2MASS flux densities. The SDSS dereddened spectrum is shown
  in blue. Gray lines represent two SWIRE templates of a 13 Gyr old elliptical
  galaxy normalized to the W2 band (solid) or to the SDSS spectrum
  (dotted). The departure of the SDSS spectrum from the latter template
  highlights the reduced Dn(4000) value of the source, while the W3 (and W4)
  excess appears from the comparison of the {\it WISE} data with the first
  template. Bottom: rest-frame SDSS spectrum with indication of the main AGN
  emission lines (upper labels) and galactic absorption lines (lower labels).}
\label{scelta}
\end{figure}

\begin{table*}
\caption{BL Lac objects and candidates in the BH12-A sample at $z \le 0.1$}
\label{bzall}
\centering                   
\begin{tabular}{rrrcccc}
\hline\hline
ID & R.A. & Dec. & $z$ & $F_{\rm 1.4 GHz}$ & log $L_{\rm radio}$ & BZCAT\\
\hline
\multicolumn{7}{c}{BH12-A sample with $z \le 0.1$}\\
\hline
 507  &   118.654457 &  39.17994 & 0.096   &  48.80 & 40.20 & BZBJ0754+3910\\
2061  &   119.695808 &  27.08766 & 0.099   &  63.50 & 40.34 & BZBJ0758+2705\\
1906  &   122.412010 &  34.92701 & 0.082   &  154.9 & 40.56 & BZBJ0809+3455\\
5997  &   127.270111 &  17.90440 & 0.089   &  231.8 & 40.81 & BZBJ0829+1754\\
 786  &   193.445877 &   3.44177 & 0.066   &  78.50 & 40.06 & BZBJ1253+0326\\
4156  &   196.580185 &  11.22771 & 0.086   &  117.9 & 40.48 & -            \\ 
4601  &   199.005905 &   8.58712 & 0.051   &   18.1 & 39.19 & -            \\ 
2440  &   212.955994 &  52.81670 & 0.076   &  321.6 & 40.81 & -            \\  
4022  &   233.009293 &  30.27470 & 0.065   &  61.20 & 39.95 & BZBJ1532+3016\\
\hline
\multicolumn{7}{c}{BH12-A sample with $0.1 < z \le 0.15$}\\
\hline
 1315  & 14.083661  &$-$9.60826   & 0.103   &   99.3 & 40.58 & BZBJ0056$-$0936 \\ 
 1805  & 119.060333 &  27.40993   & 0.140     &   20.2 & 40.17 & - \\
 1903  & 121.805740 &  34.49784   & 0.139     &   32.2 & 40.36 & - \\ 
 2083  & 132.650848 &  34.92296   & 0.145   &   31.2 & 40.39 & BZBJ0850+3455  \\  
 2281  & 144.188004 &   5.15749   & 0.131     &   30.3 & 40.29 & BZUJ0936+0509 \\ 
 2341  & 151.793533 &  50.39902   & 0.133     &   30.6 & 40.30 & BZBJ1007+5023 \\
 6152  & 162.411652 &  27.70362   & 0.144     &   15.6 & 40.08 & -             \\ 
 1845  & 163.433868 &  49.49889   & 0.140   &   56.2 & 40.62 & BZBJ1053+4929   \\ 
 6428  & 169.276047 &  20.23538   & 0.138   &  117.5 & 40.92 & BZBJ1117+2014   \\ 
 3591  & 172.926163 &  47.00241   & 0.126     &  127.4 & 40.87 & -             \\ 
 3951  & 183.795746 &   7.53463   & 0.136   &   81.7 & 40.75 & BZBJ1215+0732   \\ 
 3958  & 185.383606 &   8.36228   & 0.132   &  150.1 & 40.98 & BZBJ1221+0821   \\ 
 6943  & 194.383087 &  24.21118   & 0.140   &   10.6 & 39.89 & BZBJ1257+2412   \\ 
 1768  & 201.445267 &   5.41502   & 0.135     &   16.6 & 40.05 & -             \\ 
 7186  & 211.548355 &  22.31630   & 0.128     &    8.3 & 39.70 & - \\
 6982  & 212.616913 & 14.64450    & 0.144   &  434.4 & 41.53 & -               \\ 
 3091  & 216.876160 &  54.15659   & 0.106     &   28.5 & 40.06 & BZBJ1427+5409 \\ 
 3403  & 217.135864 &  42.67252   & 0.129   &   43.6 & 40.43 & BZBJ1428+4240   \\ 
 7223  & 223.784271 &  19.33760   & 0.115     &   11.2 & 39.73 & -  \\                                                    
 4756  & 229.690521 &   6.23225   & 0.102   &  210.9 & 40.89 & -               \\ 
 1087  & 234.945602 &   3.47208   & 0.131     &   11.3 & 39.86 & - \\
\hline
\multicolumn{7}{c}{BH12-B sample with $z \le 0.15$}\\
\hline
  1103 & 164.657227 &  56.469769 &   0.143 &   221.4 &   41.23 & BZBJ1058+5628\\ 
  1936 & 176.270935 &  19.606350 &   0.022 &   545.4 &   39.91 & - \\
  5076 & 180.764603 &  60.521980 &   0.065 &   171.4 &   40.39 & BZBJ1203+6031\\
  9591 & 184.054611 &  49.380711 &   0.146 &     9.5 &   39.88 & - \\
  9640 & 227.671326 &  33.584648 &   0.114 &     4.5 &   39.33 & BZBJ1510+3335\\
 10537 & 233.696716 &  37.265148 &   0.143 &    22.1 &   40.23 & BZBJ1534+3715\\
\hline\hline
\end{tabular}

Column description: 1) ID number, 2) R.A. [deg], 3) Dec. [deg], 4) redshift,
5) flux density at 1.4 GHz of the FIRST central component [mJy], 6) the
corresponding luminosity [$\ergs$] having adopted a null spectral index
  for the k-correction, 7) BZCAT name.
\end{table*}

As an example in Fig.\ \ref{scelta} we show a SED and SDSS spectrum of one of
the selected objects, namely the BL Lac candidate ID=2440 at $z=0.076$. The
departure of the SDSS spectrum and {\it WISE} data from the properly normalized
galactic templates highlights the reduced Dn(4000) value of the source, as well
as the W3 (and W4) excess, we interpret as due to the presence of the jet
emission.

\subsection{Expanding the search to $0.1 < z \le 0.15$}

\begin{figure*}
   \includegraphics[width=95mm]{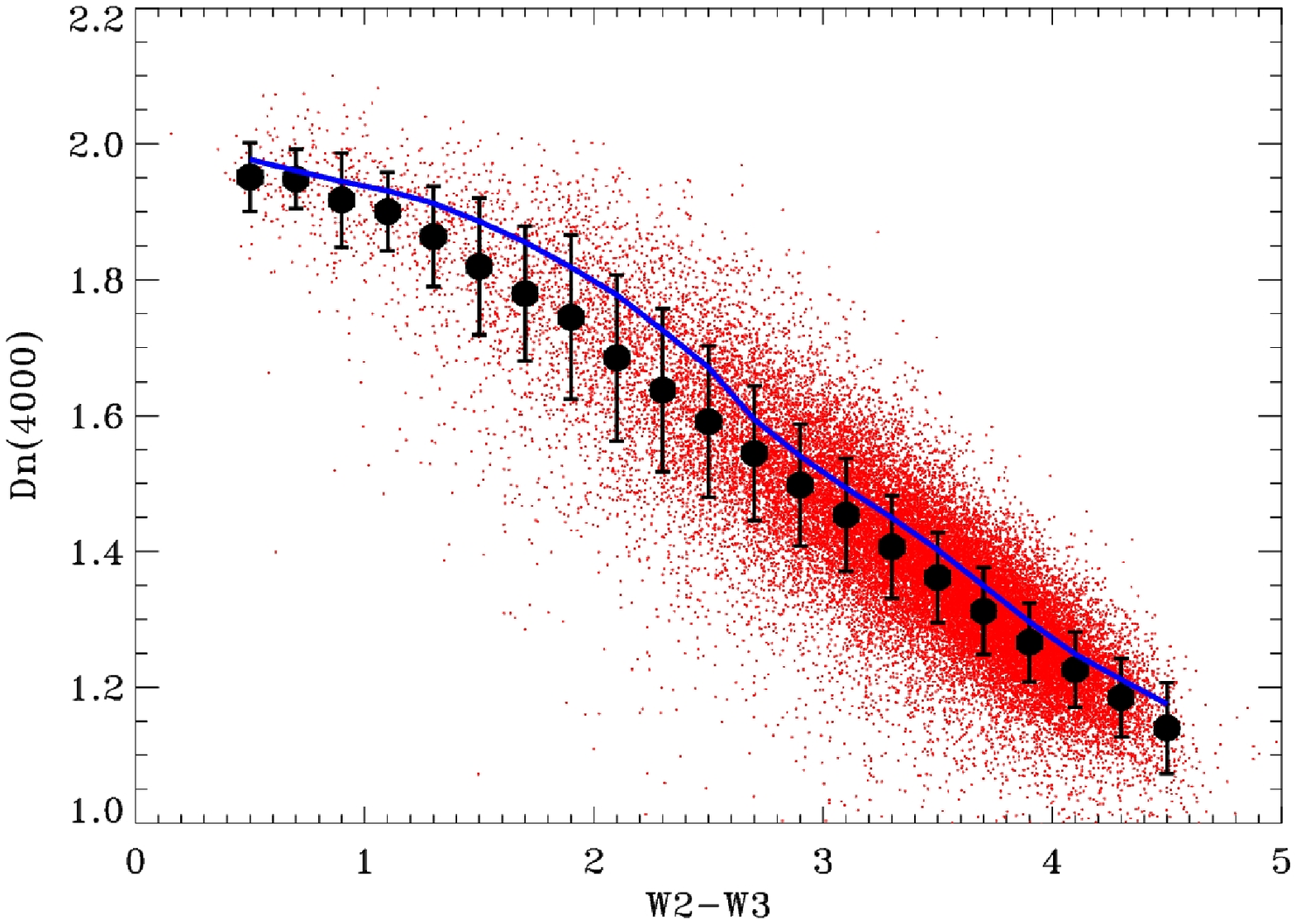}
   \includegraphics[width=95mm]{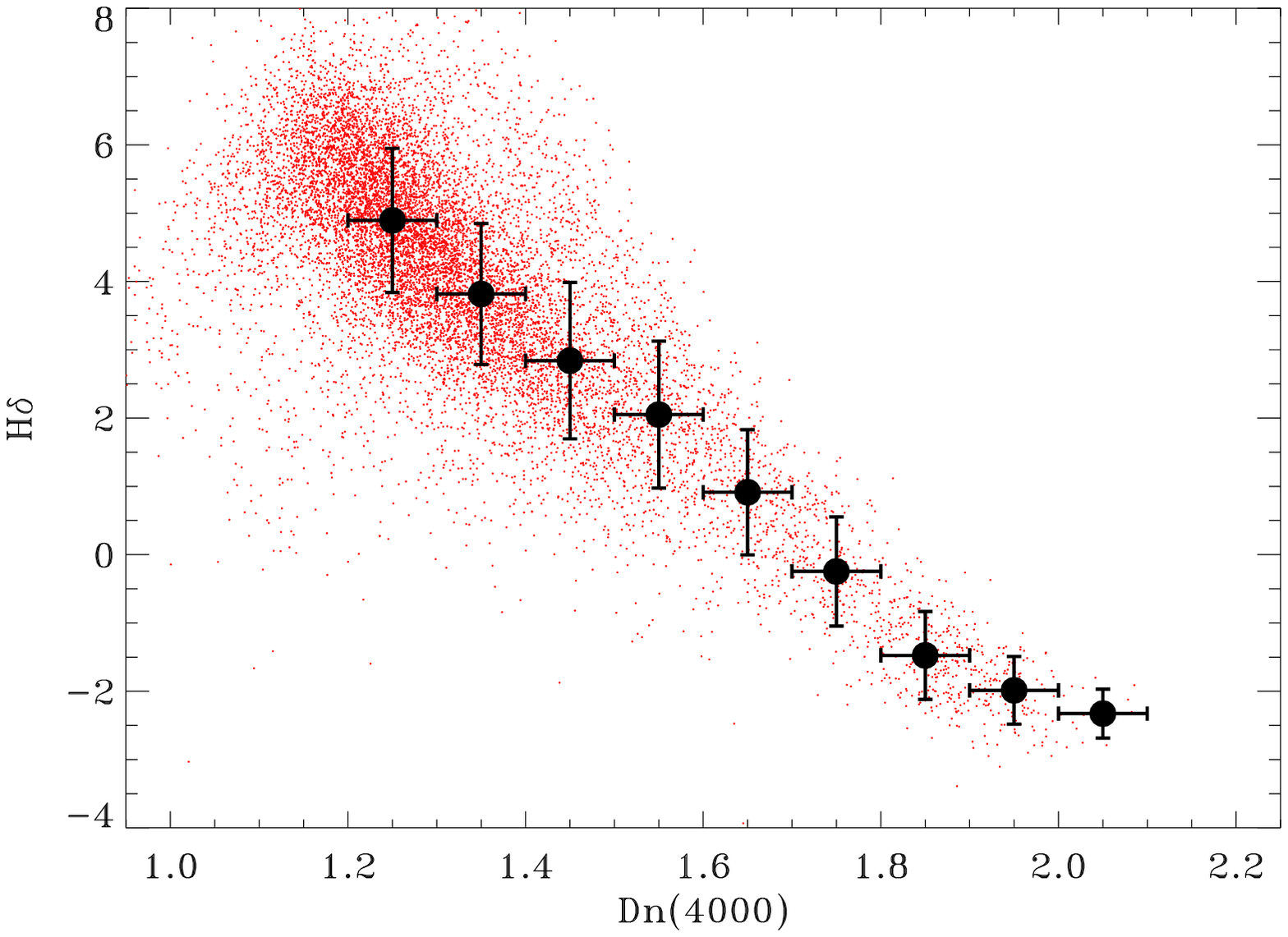}
    \caption{Same as Fig.\ \ref{uvmir} (left panel) and Fig.\ \ref{hddn} (left panel), but for $0.1 < z \le 0.15$.} 
    \label{all015}
   \end{figure*}

The BL Lac candidates we found in the $z \le 0.1$ range have all radio flux densities greater than 18 mJy, which is well above the limit of the BH12 sample, 5 mJy. This means that we can detect similar objects at slightly higher redshifts. As a result, to improve the statistical significance of our work, we expand our search for BL Lac candidates to a three times larger space volume, i.e.\ to $z=0.15$. 
We follow the same method as presented in the previous section, but we recalibrate all relationships with objects in the proper redshift range (see Fig.\ \ref{all015}).

\begin{figure*}
\sidecaption
\includegraphics[width=12cm]{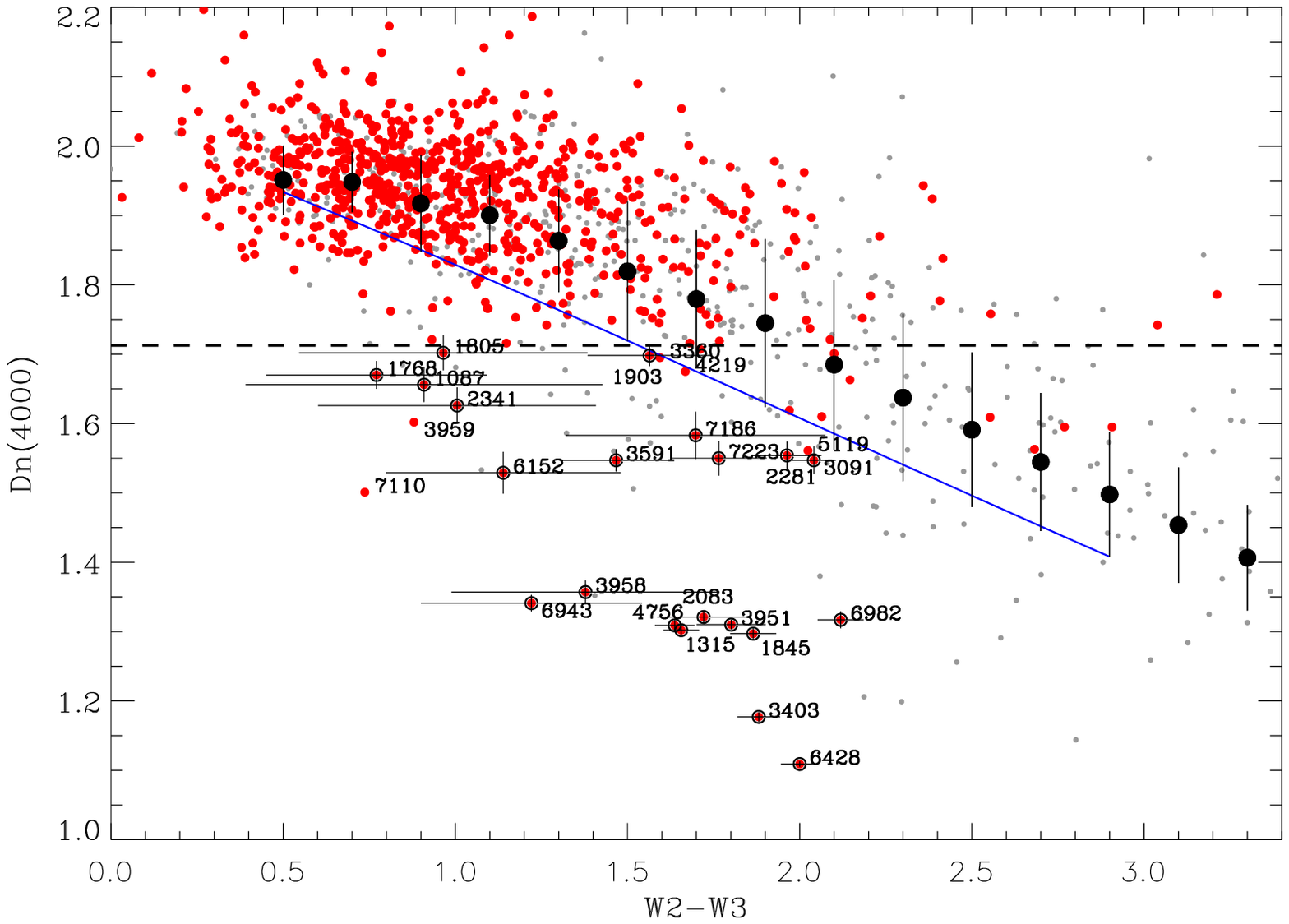}
\caption{Same as Fig.\ \ref{dn_wise_z1}, but for $0.1 < z \le 0.15$.}
\label{dn_wise_z2}
\end{figure*}

\begin{figure}
\resizebox{\hsize}{!}{\includegraphics{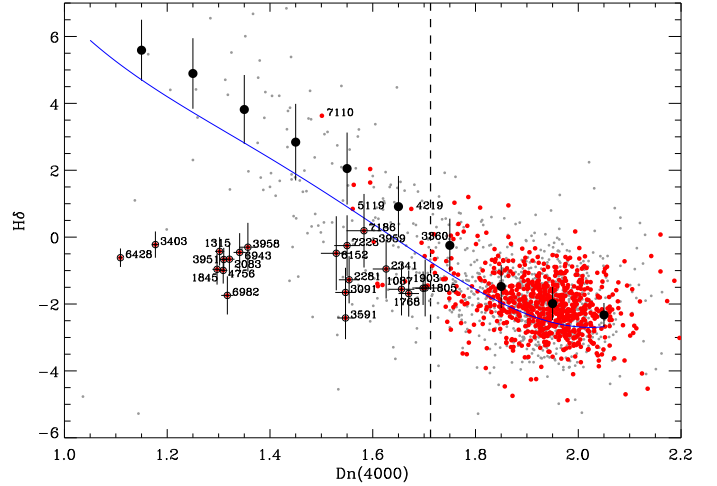}}  
\caption{Same as Fig.\ \ref{best_lick_z1}, but for $0.1 < z \le 0.15$.}
\label{best_lick_z2}
\end{figure}

In the redshift range $0.1 < z \le 0.15$, the Dn(4000) limit that corresponds to
$f=1/3$ is 1.713, since the average Dn(4000) value of quiescent galaxies
  in this redshift bin is reduced to 1.95. Our selection criteria identify
742 objects with weak emission lines.  The Dn(4000) versus W2$-$W3 diagram
(Fig.\ \ref{dn_wise_z2}) is very similar to the one obtained at lower redshift,
but there are more sources with weak lines in the ``forbidden zone".  In
particular, there are ten objects with low Dn(4000), below 1.4, and well
separated from the other sources.  The same objects stand alone in the $\rm
H\delta$ versus Dn(4000) plot shown in Fig.\ \ref{best_lick_z2}.

Moreover, there are 16 other objects that are below the 1-$\sigma$ limit of the mean Dn(4000) versus W2$-$W3 relationship.
Twelve of them are also below the 1-$\sigma$ limit of the mean $\rm H\delta$ versus Dn(4000) relationship. 
The SDSS image of the source with ID 3959 shows a contaminating star, so we discard it. 
The remaining 11 sources are BL Lac candidates. 

The properties of all candidates are reported in Table \ref{bzall}. 

\subsection{Completing the analysis of the BH12 sample}
\label{noagn}

\begin{figure*}
\sidecaption
\includegraphics[width=12cm]{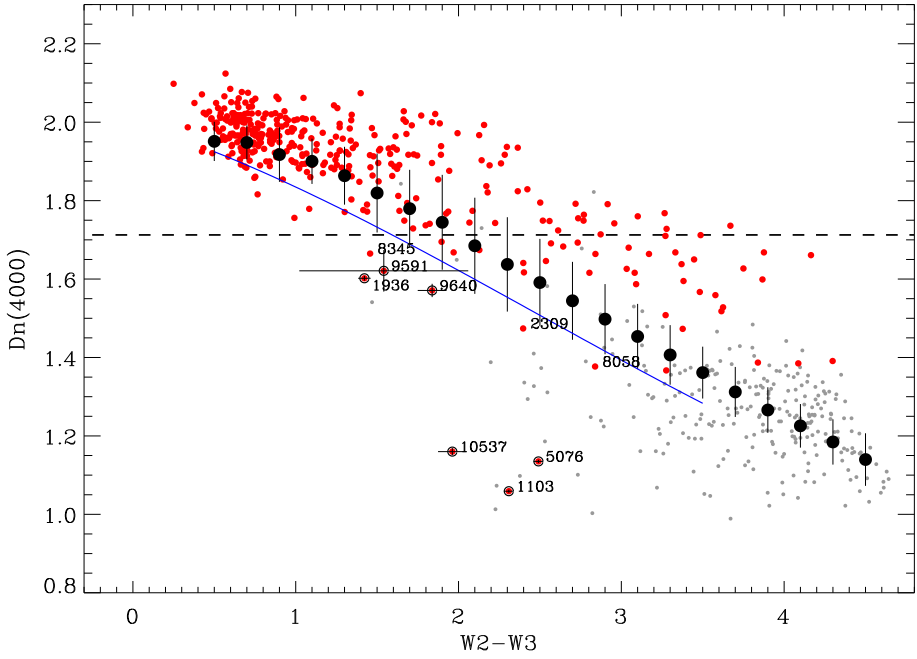}
\caption{Same as Fig.\ \ref{dn_wise_z1}, but for the BH12-B sample at redshift $z \le 0.15$. }
\label{dnw_noagn}
\end{figure*}

So far we only considered the 7302 objects included in the BH12-A radio-AGN sample, i.e.\ sources from the main galaxy sample where we excluded star-forming galaxies.
We checked whether some BL Lacs could hide among the remaining BH12 10~984 objects (hereafter BH12-B sample), 10~747 of which have a {\it WISE} counterpart.
If we consider all the sources with $z \le 0.15$ that meet the selection criteria discussed above,
we are left with 377 objects.
These reduce to 13 if we require that the position in the Dn(4000) versus W2$-$W3 is below the average relation defined for the MPA-JHU sample. However, one source 
is contaminated by a close star, two objects 
are QSO with mismatched redshift, one 
has an extremely uncertain redshift. 
Figure \ref{dnw_noagn} shows the Dn(4000) versus W2$-$W3 diagram for the BH12-B sample. The remaining nine sources are indicated. When considering that three of them (ID 2309, 8058, 8345) have high H$\delta$ index and are thus likely E+A galaxy contaminants, we are left with six BL Lac candidates. They are listed in Table \ref{bzall}. 

\bigskip
We stress that the constraint based on the W2$-$W3 color allowed us to remove
54 objects with rest-frame EW of the main emission lines $< 5 \AA$ that
would instead be selected as candidates based on the Dn(4000) index alone.
Their location in the Dn(4000) versus W2$-$W3 diagram suggests that these
objects are low-luminosity LINERs and/or Seyfert galaxies (see
Fig.\ \ref{uvmir}). Including the MIR data thus substantially improves
the purity of the resulting BL Lac sample.

\subsection{Properties of the selected objects}
\label{candies}
Our method allowed us to isolate 36 BL Lac objects (or candidates).  A
detailed analysis of each individual source is deferred to a future
paper. Nonetheless, we note that among the 15 new BL Lac candidates, ten are
associated with an X-ray source according to the ASDC SED builder
tool\footnote{\tt http://tools.asdc.asi.it/SED/}, supporting their
identification as genuine BL Lacs.  Moreover, all but one object are massive
elliptical galaxies, with stellar velocity dispersion $\sigma > 160 \, \rm km
\, s^{-1}$, which means $\log (M_{\rm BH} / M_\odot) > 7.8$ \citep{tre02},
i.e.\ the typical hosts of radio-loud AGNs. Their radio morphologies are
consistent with those characterizing BL~Lacs \citep{ant85}; i.e., they are
dominated by a compact radio core in the FIRST images, and in some cases,
    they also show a one-sided jet or a halo.

\begin{figure}
\resizebox{\hsize}{!}{\includegraphics{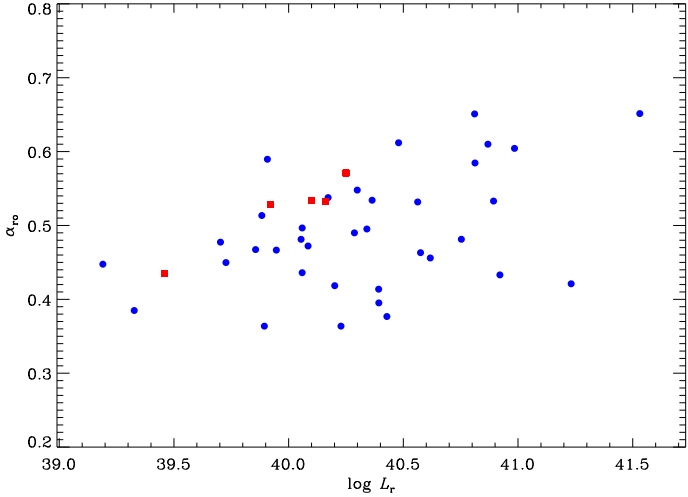}}  
\caption{Radio--optical spectral index $\alpha_{\rm ro}$ versus the
  logarithm of the radio luminosity $L_r$ for the 36 BL Lac candidates in the
  ``forbidden" zone of the Dn(4000) versus $\rm W2-W3$ diagram (blue
  dots). The red squares correspond to the five objects located in the
    ``wedge" region of the same diagram.}
\label{alfa_lr_tot}
\end{figure}

We derive the radio--optical\footnote{Estimated between 1.4 GHz and 3900 \AA,
  both in the source rest-frame; we adopted a null radio spectral index and no
  k-correction is applied.} spectral index $\alpha_{\rm ro}$ of the
  non-thermal emission for our candidates by estimating the jet optical
  contribution from the observed Dn(4000) value; i.e.
$$F_{\rm j,o} = {\rm {D_n^0-Dn(4000)} \over {D_n^0-1}} \, F_{\rm o},$$ where
  $F_{\rm o}$ is the SDSS flux density (reddening-corrected) at 3900 \AA.
  Figure \ref{alfa_lr_tot} shows $\alpha_{\rm ro}$ versus the logarithm of the
  radio luminosity of the FIRST central component $L_{\rm r}$.  The mean value
  is $<\alpha_{\rm ro}> = 0.49$ and the standard deviation 0.08. The range of
  $\alpha_{\rm ro}$ covered by our objects overlaps that of the BL Lacs
    analyzed by \citet{sambruna96}, \citet{pad03}, and \citet{donato01}, going from $\sim 0.2$ to
  $\sim$ 0.75.

\begin{figure}
\includegraphics[width=9cm,angle=0]{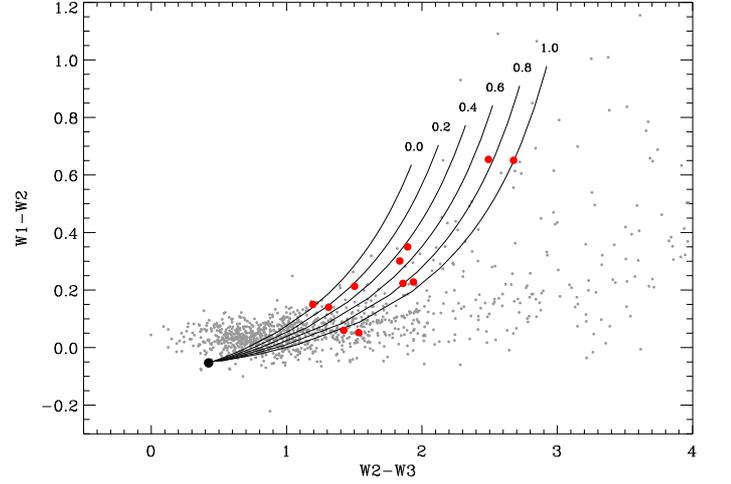}
\caption{{\it WISE} color-color diagram for the BH12 sample with S/N(W3) $>$ 2 and $z \le 0.1$ (gray dots). Red filled circles represent our BL Lac candidates in the same redshift range.
The black tracks indicate the sequence followed by objects where a power-law component $F_\nu \sim \nu^{-\alpha}$ of increasing strength is added to the emission of an elliptical galaxy, represented by the black dot. This was derived from the SWIRE template of a 13 Gyr elliptical galaxy at $z=0.1$. Different tracks correspond to different values of $\alpha$ in the range 0--1.}
\label{wise_tot}
\end{figure}

Figure \ref{wise_tot} shows the location of our BL Lac candidates in the {\it WISE} color-color diagram.
They follow a sequence, starting from the region where passive elliptical galaxies are located \citep[e.g.,][]{wri10,rai14}, and moving toward the so-called ``{\it WISE} blazar strip" identified by \citet{masf11}. We also show tracks corresponding to the addition of a power-law emission component $F_\nu \sim \nu^{-\alpha}$ of increasing strength to a $z=0.1$ elliptical galaxy flux. These tracks were obtained starting from the 13 Gyr elliptical galaxy template of the SWIRE Template Library\footnote{http://www.iasf-milano.inaf.it/$\sim$polletta/templates/swire\_templates.html} \citep{pol07} for different values of $\alpha$ in the range 0--1. They cover the region where most of our candidates are found. This supports the presence of a power-law emission component in the MIR bands, whose relative weight increases at longer wavelengths.

Overall, the properties of our candidates are reassuringly consistent with those of classical BL Lacs.

\section{Completeness of the low-luminosity BL Lacs sample}

\begin{figure}
\includegraphics[width=9cm,angle=0]{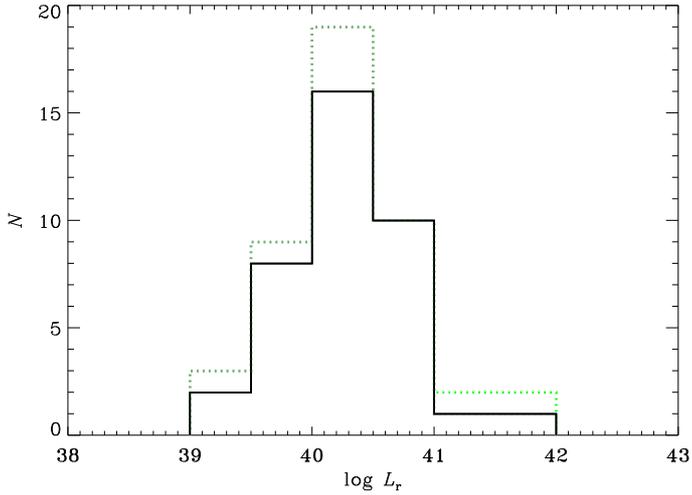}
\caption{Radio luminosity distribution at 1.4 GHz of the 36 BL Lac
  candidates. The dotted black histogram includes the five sources in the
  ``wedge'', while the green one also considers the two BL Lacs tagged as
  ``STAR'' in the SDSS, see Sects.\ \ref{wedges} and \ref{bzcat}.}
\label{histol}
\end{figure}

Figure \ref{histol} shows the radio luminosity distribution of the FIRST
central component, $L_{\rm r}$, for the 36 BL Lacs identified with our method.
They extend up to $\log L_{\rm r} \sim 41.5 \, \ergs$.  The number of BL Lac
candidates rapidly increases toward lower $L_{\rm r}$, reaching a peak at
$\log L_{\rm r} \sim 40.5 \, \ergs$ and then falls.  This drop can be due to a
genuine paucity of low-luminosity objects or, alternatively, to a decrease in
the completeness of our sample. Before deriving the BL Lac radio luminosity
function (RLF), we must consider the selection effects included in the various
steps described in Section \ref{sase}.

\subsection{Completeness of the NVSS-SDSS sample}

From the point of view of the radio completeness, BH12 included the sources
with a NVSS flux density larger than 5 mJy. The vast majority of our selected
BL Lac candidates have fluxes that are much larger than this limit, an indication that
the flux threshold does not have a significant impact on our
results. Nonetheless, we include the radio flux limit in the simulations
presented in Sect.\ \ref{simu}.

\begin{figure}
\resizebox{\hsize}{!}{\includegraphics{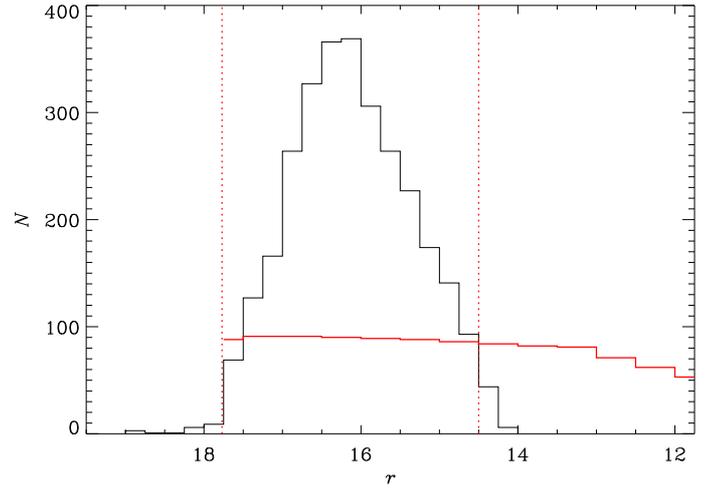}}  
\caption{Distribution of $r$ band magnitudes of the galaxies in the BH12-A
  sample with $0.05 \le z < 0.15$. The vertical dotted lines indicate the limits
  defining the SDSS main galaxies sample. The red histogram reports the SDSS
  completeness in percentage from \citet{mon09}.}
\label{complete}
\end{figure}

As for the optical selection of the sample, according to \citet{mon09} the
redshift completeness of the SDSS decreases with decreasing apparent
magnitude, starting from $\sim 90\%$ at the SDSS spectroscopic limit of
$r=17.77$ and reaching $\sim$50\% at $r=11.75$ (see Fig. \ref{complete}). Most
of the incompleteness is due to the SDSS fiber cladding that prevents fibers on
any given plate from being placed closer than 55\arcsec\ apart. This does not
introduce a bias into the sample, but leads to a random loss of $\sim 10\%$
\citep{zeh02} of the potential spectroscopic targets.  For brighter (and more
extended) objects, other effects become important, such as the superposition of
bright saturated stars on the target.

Although the radio-loud AGN (RL-AGN) hosts are very luminous galaxies, it is
important to assess their completeness level from the point of view of the
SDSS spectroscopy. In Fig. \ref{complete} we show the distribution of $r$ band
magnitudes of the galaxies in the BH12-A sample with $0.05 \le z < 0.15$. The
distribution peaks at $r\sim16.2$, and it rapidly falls on both sides. The
number of objects is essentially reduced to zero before reaching the high
and low magnitude limits imposed by their selection in the SDSS main galaxies
sample ($r$=17.77 and 14.5, respectively). This indicates that the selected
sample of galaxies is complete.

Conversely, very nearby (and bright) galaxies might fall in the region of low
completeness at low magnitudes. We then prefer to set a more stringent lower
limit of $z=0.05$ on the galaxies considered. From the point of view of the BL
Lac candidates, we discard only one of them (ID~1936), while from the point of
view of the volume sampled, this is reduced by just 4\%.

\subsection{Check on the {\it WISE} constraints}

Could our requirements on the existence of a {\it WISE} counterpart and/or on the quality of {\it WISE} data affect the BL Lac candidates count?
Among the 60 sources in the BH12 AGN sample without a {\it WISE} counterpart, there are no objects meeting our criteria to be considered BL Lac candidates: EW less than 5 \AA, existence of a FIRST counterpart, and $\rm Dn \le Dn_{\rm max}$.

If we consider the 445 sources with S/N(W3)$\le 2$ in the BH12-A sample, only
three objects meet the selection criteria. However, the low quality of the
overall multiband dataset for these sources suggests that we should not include them among
the BL Lac candidates. In fact, not only are the {\it WISE} data very uncertain,
but the optical indices also have large errors. We verified that among the
237 objects in the BH12-B sample without a {\it WISE} counterpart or with low
S/N(W3), there are no BL Lac candidates.

\subsection{Are there BL~Lacs outside the forbidden zone?}
\label{wedges}

The selection method we propose is based on the idea that the BL~Lacs are
  found in the ``forbidden'' zone, while other radio emitting galaxies are
  located along a well-defined locus in the Dn(4000) vs. W2$-$W3 plane.
  However, there is the possibility that some BL~Lacs follow this same
  sequence. Indeed, the simulated tracks presented in the right hand panel of
  Fig. \ref{uvmir} indicate that the objects with a spectral index in the MIR -
  optical bands $\gtrsim$ 1 overlap with this region.  

We then consider those sources that meet the same requirements as the ones used to select
BL~Lacs objects (rest-frame EW of the emission lines $<$ 5
\AA\ and existence of a FIRST central counterpart) and that are located in the
``wedge" region of the Dn(4000) vs. W2$-$W3 plane.  This is the region below
the Dn(4000) limit but above the polynomial fit to the $1-\sigma$ limit of the
relationship derived for the MPA-JHU sample (see Fig.\ \ref{dn_wise_z1}). We
find 54 sources, 22 of which have $0.05 \le z < 0.15$. When discarding the E+A
galaxies, their number is reduced to ten. Five of them are spiral galaxies, not
the typical BL~Lac hosts, and they are most likely low-luminosity AGNs.  The
remaining five are elliptical galaxies: while one is a BZCAT object (probably
a genuine BL Lac), the nature of the other four is highly
uncertain. Their compact radio morphology does not provide us with any useful
information. We list these objects in Table \ref{wedge}.  They have Dn(4000)
values of 1.6--1.7 and $\rm W2-W3$ colors in the range 1.7--2.1; i.e., they
are all concentrated in a small portion of the ``wedge" region, close to its
apex. Their radio luminosity ranges from $\log L_{\rm r} \sim 39.5$ to $40.3
\, [\ergs$], see Fig.\ \ref{histol}, and their $\alpha_{\rm ro}$ goes from
0.44 to 0.57 (see Fig.\ \ref{alfa_lr_tot}).

\begin{table*}
\caption[]{Possible BL~Lacs in the BH12 sample with $0.05< z \le 0.15$ in the
  ``wedge" region of the Dn versus $\rm W2-W3$ diagram.}
\label{wedge}
\centering
\begin{tabular}{rrrcccccc}
\hline\hline 
  ID & R.A.      & Dec.       & $\rm W2-W3$ & Dn(4000)  & $z$   & $F_{\rm 1.4 GHz}$ & log $L_{\rm radio}$ & BZCAT\\
\hline
 935 & 147.13342 & 55.59314   & 2.064       & 1.610 & 0.118 & 28.6 & 40.16 & BZUJ0948+5535\\
3062 & 194.97218 & 57.86380   & 1.711       & 1.707 & 0.149 &  9.9 & 39.92 & -\\
5525 & 213.01450 & 29.46714   & 1.970       & 1.619 & 0.115 & 37.0 & 40.25 & -\\
2468 & 226.09259 & 47.68667   & 1.957       & 1.721 & 0.093 &  9.5 & 39.46 & -\\
1278 & 359.67532 & $-9.90999$ & 2.146       & 1.663 & 0.105 & 31.8 & 40.10 & -\\
 \hline\hline
\end{tabular}

Column description: 1) ID number, 2) R.A. [deg], 3) Dec. [deg], 4) W2$-$W3,
5) Dn(4000), 6) redshift, 7) flux density at 1.4 GHz of the FIRST central
component [mJy], 8) the corresponding luminosity [$\ergs$], 9) BZCAT name.
\end{table*}

\subsection{An external completeness test: the BZCAT}
\label{bzcat}
We here assess the completeness of our sample against the compilation
of BL Lac objects (or candidates) included in the multifrequency catalog of blazars Roma-BZCAT.
The edition 4.1.1 (August 2012) contains 3149 blazars, 1221 of which are BL Lac objects or candidates (BZB),  and 221 are blazars of uncertain type (BZU).
Such objects have been selected with different methods and from observations in various bands. The comparison might unveil the presence of bias in our selection method.
Details on the cross-match between the BZCAT sources and our sample are given in Appendix \ref{abzcat}.

In summary, in the $0.05 < z \le 0.1$ redshift range, the comparison between
the BH12 sample and BZCAT suggests that we did not miss any object because of our
selection method.  We only lost 3 objects out of 18 because of fiber
collision, which does not, however, introduce any bias into the sample but only a
lower effective sky coverage.

Moving to higher redshift, we recover all BZB and lineless BZU with the
exception of six objects: two are bright featureless sources that are not
included in the BH12 sample because they are tagged as ``STAR". We 
  consider them in the RLF discussion. One object was instead missed because
  it lies in the ``wedge" region of our Dn(4000) versus W2$-$W3 diagnostic
  plane. As shown in Fig.\ \ref{uvmir}, this is the region populated by
  sources with MIR-optical spectral index $\alpha \sim 1$, so likely LBL
  objects. Finally, three were discarded by our method because of a high
  Dn(4000) value, between 1.95 and 2.01. However, two of them also have {\it WISE}
  colors that are typical of quiescent galaxies, W2$-$W3 = 0.76 and 0.39,
  respectively. For one of them, as detailed in the Appendix, we have strong
  evidence that it might be an incorrect identification.

\begin{figure*}
\includegraphics[width=9cm,angle=0]{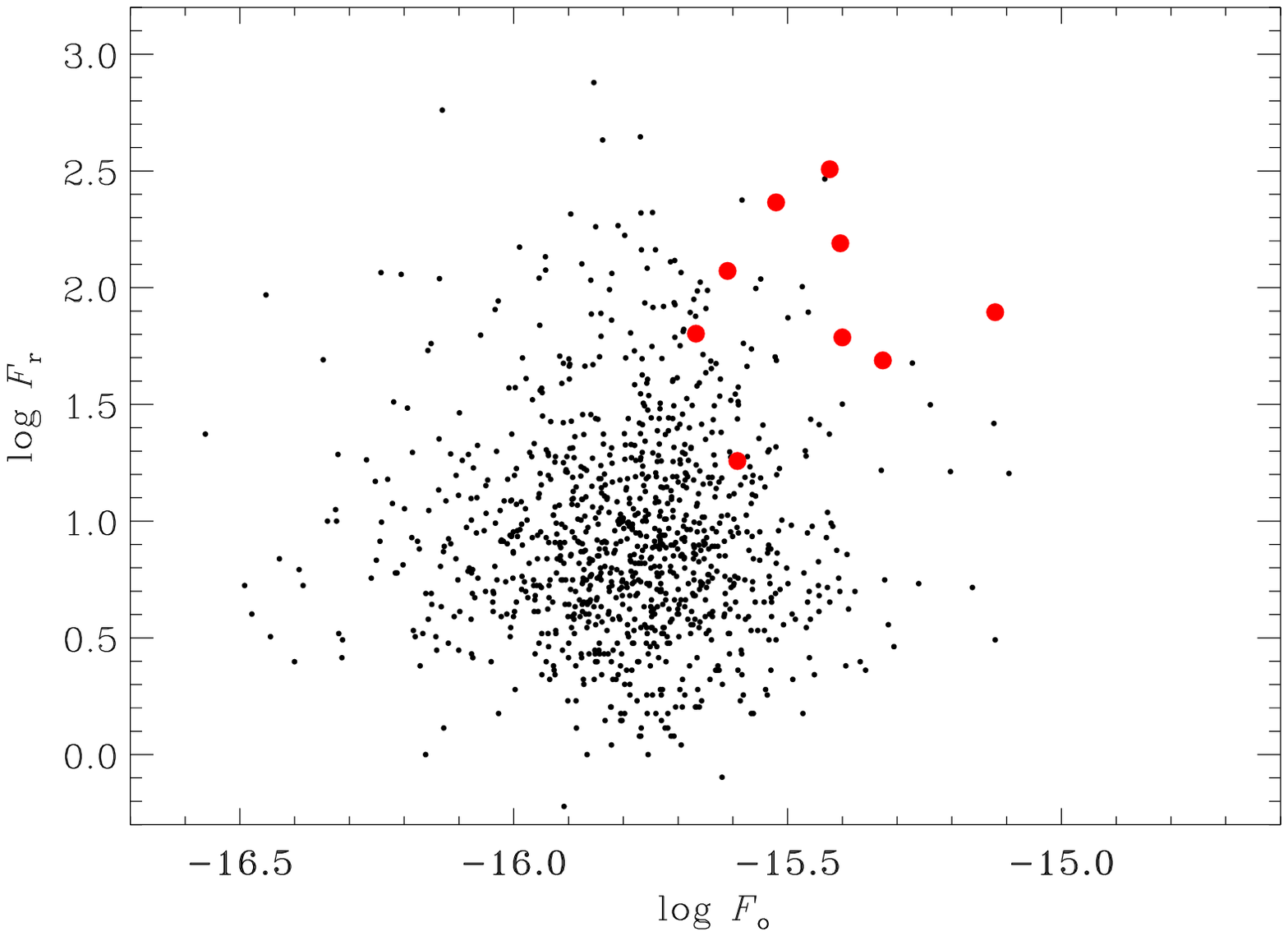}
\includegraphics[width=9cm,angle=0]{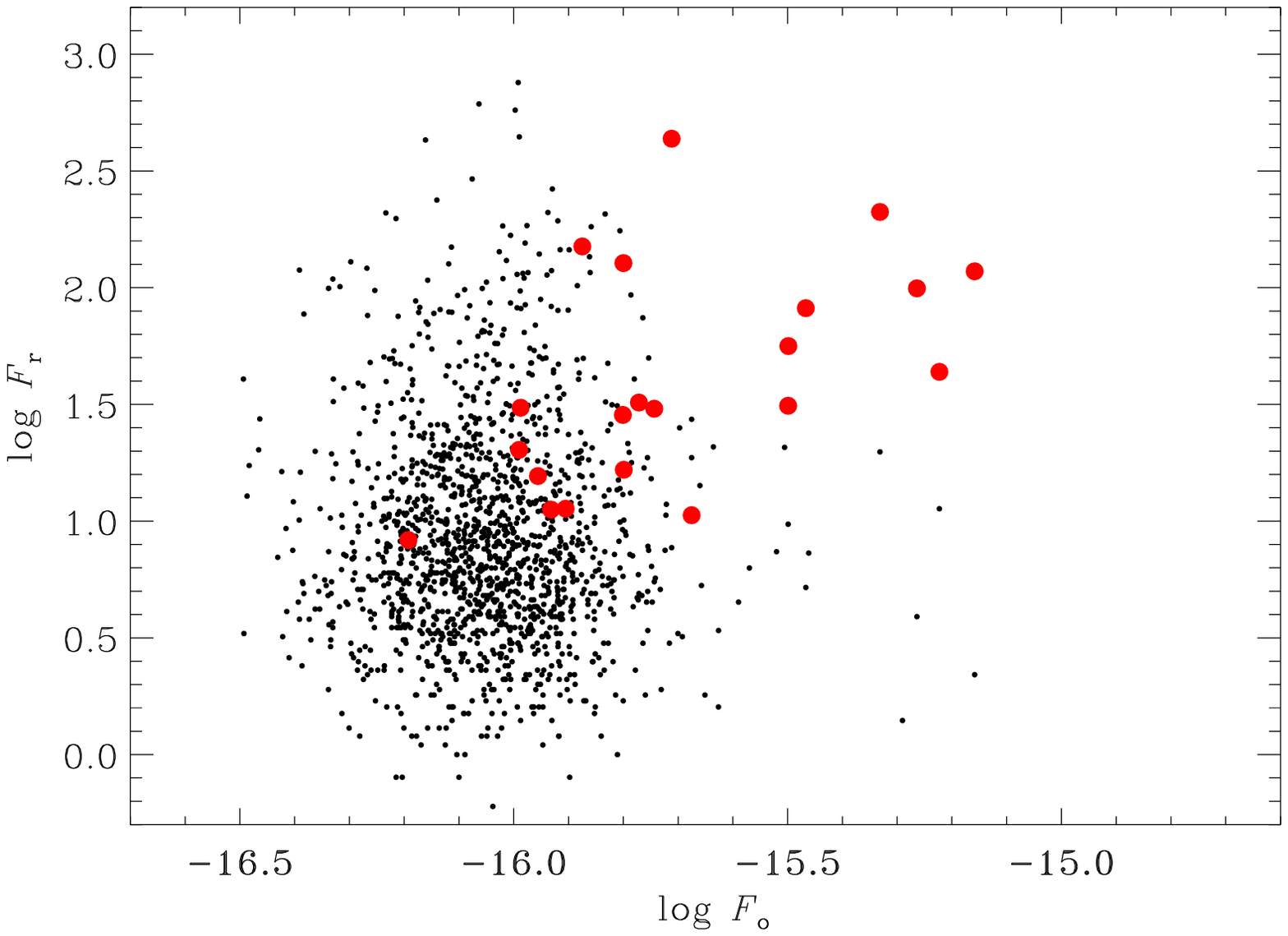}
\caption{Left panel: the flux density of the FIRST central component, $F_{\rm
    r}$, versus dereddened flux density within the SDSS fiber at 3900 \AA,
  $F_{\rm o}$, for the sources with $0.05 < z \le 0.1$. The candidate BL Lacs
  are plotted with red circles. Right panel: same as in the left panel, but
  for $0.1 < z \le 0.15$. }
\label{detect}
\end{figure*}

\section{BL Lac selection function}
\label{simu}

As discussed in Sect.\ \ref{sase}, the main requirement for including
an object belonging to the NVSS-SDSS sample in our candidates list is a flux
contribution from the non-thermal jet component, $F_{\rm{j,o}}$, of at least
one-third of the galactic one at 3900 \AA\ (rest frame).

To estimate how this criterion is reflected in the selection
function at different values of radio luminosity, we ran a Monte Carlo
simulation. In each $L_{\rm r}$ bin, we extracted 100~000 objects with:
\begin{itemize}
\item[i)] a random distribution in redshift between $0.05 < z \le 0.15$,
  properly weighting for the appropriate volumes and assuming a constant
    source density with redshift, to derive the corresponding radio flux;
  objects with radio flux density below the 5 mJy threshold of the BH12 sample
  are discarded;
\item[ii)] a random spectral index $\alpha_{\rm ro}$ (see below for a
  detailed discussion on this parameter) that leads to an estimate, for each
  object, of the expected jet optical emission $F'_{\rm{j,o}}$ [$\ergscmA$]
  as
$${\rm log} \, F'_{\rm{j,o}} = {\rm log} F_{\rm r} \, {\rm [mJy]} - 5.74 \, \alpha_{\rm ro} - 14.71 \, ;$$
\item[iii)] a random value of the flux density at 3900 \AA\ (rest frame) with
  average and dispersion measured from the sample galaxies divided in
    redshift bins of 0.01. In Fig.\ \ref{detect} we show the distribution of
  radio versus optical fluxes into two redshift ranges.
\end{itemize}

For each object we tested whether the ratio between jet and galactic optical
emission is above the $f=1/3$ threshold. The fraction of objects expected to
be selected as BL Lac candidates as a function of radio luminosity can finally
be derived.

\begin{figure}
\includegraphics[width=9cm,angle=0]{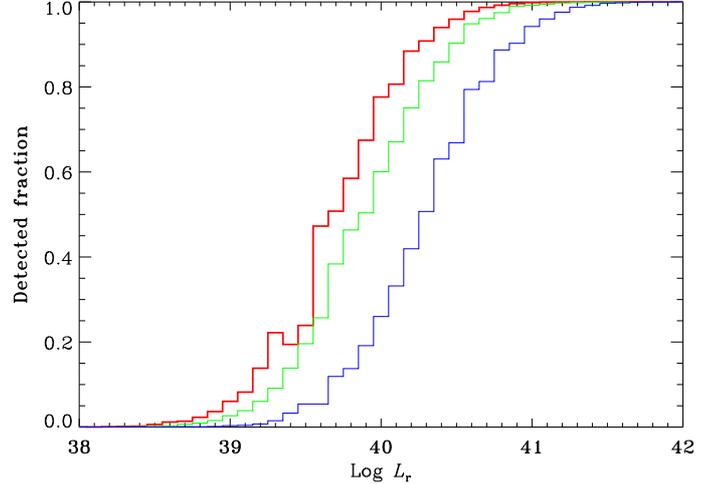}
\caption{Expected fraction of BL Lac candidates recovered with our selection
  method at various levels of radio luminosity obtained with Monte Carlo
  simulations. The various lines refer to different assumptions on the
  radio--optical spectral index. Red: $\alpha_{\rm ro}=0.49$, the mean
  obtained from the 35 BL Lac candidates with $0.05 < z \le 0.15$; green:
    $\alpha_{\rm ro}=0.53$, the mean of the candidates with $\log L_{\rm r} >
    40.5$ $\ergs$; blue: $\alpha_{\rm ro}=0.59$, the mean of the 1Jy BL~Lacs.}
\label{monte}
\end{figure}

The assumption on the distribution of the radio-optical spectral index is
  the key to this simulation. In Sect.\ \ref{candies} we estimated a mean
  value of $<\alpha_{\rm ro}> = 0.49\pm0.08$. However, our results might be
  plagued by selection effects, since high $\alpha_{\rm ro}$ objects of low
  radio luminosity could escape detection in the optical band. To
  assess how robust our estimates are, we compared them to those available in
  the literature. The average spectral index of the (radio-selected) 1Jy
  sample of BL~Lacs \citep{sti91} is $<\alpha_{\rm ro}> = 0.59$ with a
  dispersion of 0.10 \citep{sambruna96}. Adopting their average $\alpha_{\rm
    ro}$ for the simulation, we obtain the selection function reproduced as the
  blue curve in Fig. \ref{monte}: it reaches 80\% at $\log L_{\rm r} \sim
  40.5$ [$\ergs$]. This means that, at least above this threshold, the BL Lacs
  sample we selected is very close to completeness and that the distribution
  of $\alpha_{\rm ro}$ is not significantly affected by selections biases. Its
  average, $<\alpha_{\rm ro}> = 0.53 \pm 0.08$, can be considered as a robust
  value. The selection function obtained by adopting this value is shown in
  Fig.\ \ref{monte}.

The 1 Jy objects are, in general, much brighter than the BL Lacs we
are considering, with radio luminosities reaching $\sim 10^{45}$$\ergs$. If we
limit to the 1 Jy sources with $L_{\rm r} < 10^{43}$$\ergs$ we obtain
$<\alpha_{\rm ro}> = 0.52\pm$0.09, which agrees remarkably with the value
quoted above, while the average $\alpha_{\rm ro}$ for those with $L_{\rm r} >
10^{43}$ $\ergs$ is $0.65\pm$0.08. These results suggest there is a
trend between $\alpha_{\rm ro}$ and the radio luminosity, as expected from the
so-called ``blazar sequence" \citep{fos98}. This would account for the
(slightly) lower average value we find for our LPBL sample. Indeed, the
  sources of the 1 Jy sample (limiting to the 22 with measured redshift) show
  a linear correlation between $\alpha_{\rm ro}$ and $L_{\rm r}$, with a slope
  of 0.06. The Spearman rank correlation test returns a probability of only
  2\% that the two variables are not correlated. We therefore repeat the
simulation by adopting the observed distribution of spectral indices; i.e.,
$<\alpha_{\rm ro}> = 0.49$ (Fig. \ref{monte}). The selection
function obtained with this assumption differs only marginally from what is
derived for $<\alpha_{\rm ro}> = 0.53$, with a horizontal shift of only
$\sim$0.2 dex.

\section{The BL Lacs luminosity function}

\begin{figure}
\includegraphics[width=9cm,angle=0]{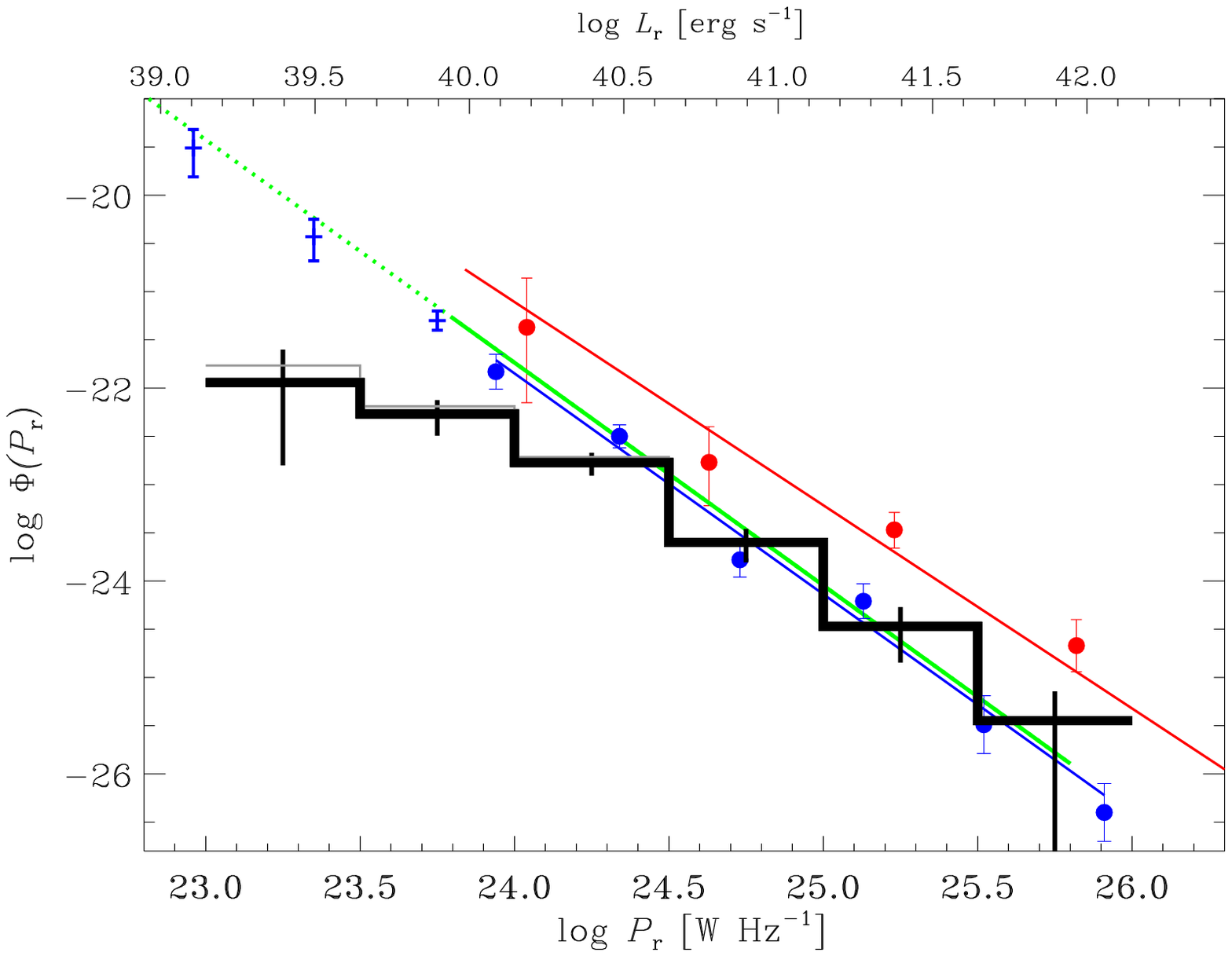}
\caption{RLF at 5 GHz of the selected BL Lac candidates (black histogram).
  The gray histogram shows the effects of including the five
    objects located in the ``wedge''. The results from \citet{pad07} are in
  red, while blue dots refer to the work by \citet{marcha2013} for ``classical
  BL Lacs''. We also report (as plus signs) three points at low luminosity of
  the \citeauthor{marcha2013} RLF for ``Type 0'' AGNs. The green line is the RLF normalized to
  match the number of objects of our sample with $\log P_{\rm r} > 23.8 \rm \,
  [W \, Hz^{-1}]$, after having adopted the same slope of \citet{marcha2013}.}
\label{lfunc}
\end{figure}

Several studies have dealt with the determination of the BL Lacs RLF.  In all cases,
the approach was based on flux-limited samples. In our study, instead, we
consider an essentially volume-limited sample, because the radio, optical, and
MIR flux limits do not represent a significant constraint in our selection.
The advantage of this method is twofold: all objects have a known redshift, and
they are all within a very limited range, so that evolution can be neglected.

We can then derive the BL Lacs RLF by simply dividing the number of detected
objects by the total volume covered, recalling that the intersection between
the spectroscopic DR7 and FIRST covers 17.3\% of the whole sky. By also considering
the effects of fiber cladding, this results in a volume of 0.158 Gpc$^3$
from $z=0.05$ to $z=0.15$. We convert the radio luminosity distribution shown
in Fig.\ \ref{histol} (including the 35 candidates with $0.05 < z \le 0.15$
and the two objects tagged as ``STAR'') by dividing the number of objects by
the radio luminosity, the selection function, and the total volume. The errors
on the RLF have been estimated by combining Poisson statistics with the
uncertainties related to the selection function\footnote{We considered the two
  assumptions: $<\alpha_{\rm ro}>$ = 0.49 and 0.53. The differences in
  the resulting RLF are of only 0.17, 0.15, and 0.05 dex for the three bins at
  the lowest luminosities. By adopting instead $<\alpha_{\rm ro}>$ = 0.59,
    these bins are further increased by 0.75, 0.38, and 0.12 dex,
    respectively.}. The result is shown in Fig.\ \ref{lfunc}. To ease the
  comparison with previous studies, we present the radio luminosities in both
  cgs and MKS units and assume a zero radio spectral index to convert from 1.4
  to 5 GHz \citep{sti91}.

We can compare our results on the RLF with those obtained by previous works
and, in particular, with \citet{pad07} and \citet{marcha2013}. We assume a RLF
in the form of a power law, $\Phi(P_{\rm r}) \propto P_{\rm r}^{-B}$.  Since
our sample covers a limited range of radio luminosities, we refrain from
calculating the RLF slope.  If we adopt a power-law slope of $B=2.31$ as in
\citet{marcha2013} for ``classical" BL Lacs, we can calculate the
normalization of our RLF so that the integrated RLF returns the number of
observed objects above $\log P_{\rm r} = 23.8 \, \rm [W \, Hz^{-1}$], the
range of powers common to both studies.  This is displayed in
Fig.\ \ref{lfunc}\footnote{Had we used $B=2.12$ as derived by \citet{pad07},
  we would have obtained a difference of only 0.08 dex at $\log P_{\rm r} =
  24.5 \, \rm [W \, Hz{^-1}$]}. The comparison shows very good
  agreement with the results obtained by \citeauthor{marcha2013} for BL Lacs
  and a mismatch between our RLF and the \citeauthor{pad07} one by a factor
  $\sim 3$.

\begin{figure}
\includegraphics[width=9cm,angle=0]{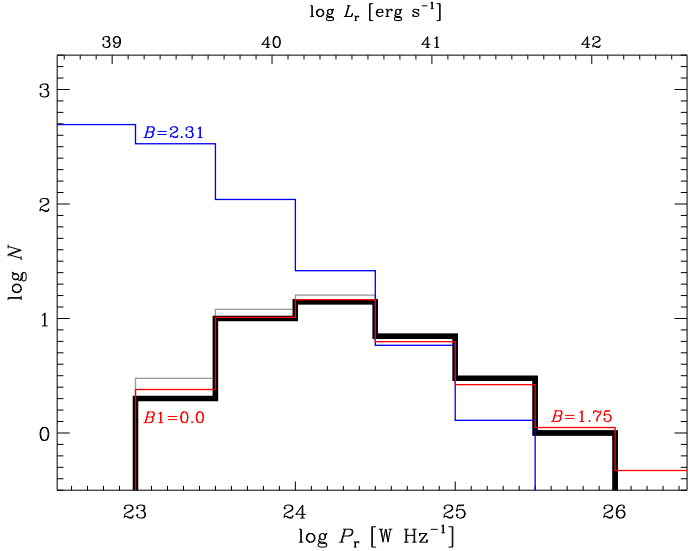}
\caption{Observed number of BL Lacs (black solid line) versus radio power. The
  blue histogram represents the predicted number of objects in our volume
  extrapolating the RLF derived for our sample to low radio luminosities
  (green line in Fig.\ \ref{lfunc}) and multiplied by the selection function
  (red line in Fig.\ \ref{monte}). The gray histogram shows the effects
    of the inclusion of the five objects located in the ``wedge''. The red
  histogram displays the same quantity as for a RLF with $B=1.75$ and with a break
  at $\log P_{\rm r} = 24.5 \rm \, [W \, Hz^{-1}]$, below which the RLF
  slope is $B1=0$.}
\label{nhist}
\end{figure}

The main aim of our study is to isolate and analyze the least luminous BL
Lacs. Indeed, we detected candidates down to $\log P_{\rm r} \sim 23$ [W
  Hz$^{-1}$]. We defer a detailed study of the individual
objects and of their SED to a future paper. Here we want to address the behavior of their
  RLF in a poorly studied regime of low radio power. The only study reaching
  these luminosities is from \citet{marcha2013}: they consider not only
  classical BL Lacs but also the objects they defined as ``Type~0'' and
  ``weak-lined'' AGNs. Their space density is substantially higher than for our
  BL~Lac sample, with, say, about ten times more Type~0 AGNs than BL~Lacs
  between $\log P_{\rm r} = 23.5$ and $ 24.0$ [W Hz$^{-1}$].  However, recent
  studies have shown that the dominant population of radio galaxies in the local
  Universe is formed by the so-called FR~0 sources
  \citep{baldi10b,sadler14,baldi15}. Their lines EW are generally low (more
  than 80\% showing values lower than 5\AA, R. Baldi, private
  communication). Moreover, they are generally sources with a high core
  dominance and, for this reason, often flat spectrum radio sources. As
  explained by \citet{baldi15} the deficit of extended radio emission is an
  intrinsic property of FR~0, and it is not due to a beaming effect of orientation, 
  as in the case of BL~Lacs. In other words, a significant
  fraction of the FR~0 population conforms to the definition of Type~0 and
  weak-lined AGNs. Therefore, the association of all Type~0 and weak-lined AGNs
  with LPBL might not be justified.

We estimate the predicted number of LPBL in our volume by extrapolating the RLF derived above 
to low powers, taking our selection function into account (see
Fig.\ \ref{monte}). The result is shown in Fig.\ \ref{nhist}, where we compare
the observed and predicted number counts. A large disagreement emerges, where the
difference is a factor of $\sim 10$ already at $\log P_{\rm r} \sim
10^{23.75} \rm \, W \, Hz^{-1}$ (i.e., $\log L_{\rm r} \sim 40$), where our
selection function is high (around 70\%) enough to make this result
robust.

 We then consider the possibility of a break in the RLF occurring at $P_{\rm
   break}$. Furthermore, it appears that also at high luminosities the slope
 might be less than the value of 2.31 adopted. We are not in the position
 of accurately determining all parameters describing the RLF with the
 available data. Nonetheless, in Fig.\ \ref{nhist} we show an illustrative
 example of a RLF with $B=1.75$, a break at $\log P_{\rm break} = 24.5 \rm \, [W
   \, Hz^{-1}]$, and a slope below the break $B1=0$, which fairly matches the
 observed counts at all luminosities.  This result is effectively
   independent of the assumption used for the radio-optical spectral index:
   had we used the average value derived for the 1 Jy sample, the break
   luminosity would have decreased by just a factor $\sim$ 1.7.

Values of $B1$ below 1, as suggested by this analysis, correspond to BL Lacs number
 counts growing with luminosity up to $P_{\rm break}$, implying the
 existence of a peak.

\subsection{The BL Lac RLF and the AGN unified model.}

According to the AGN unified scheme, the RLFs of BL Lacs and RL-AGNs are
expected to be linked to each other. 
In fact, blazars are thought to be the beamed version of radio galaxies \citep[see, e.g.,][]{urr95}. 
The effect of relativistic beaming on the RLF was analyzed in a series of papers by \citet{urr84}, \citet{urr91}, \citet{urr91a}, and \citet{urr95}. 
If the RLF of the
parent population of unbeamed objects is described by a broken power law, the
break will produce a change of slope in the BL~Lac RLF at a power amplified by $\sim \delta^3_{\rm
  max}$, where $\delta_{\rm max}$ is the maximum value
of the jet Doppler factor distribution\footnote{The Doppler factor $\delta$ is
  defined as $\delta=[\Gamma (1-\beta \, \cos \theta)]^{-1}$, where
  $\Gamma=(1-\beta^2)^{-1/2}$ is the bulk Lorentz factor, $\beta$ is the
  velocity in units of the speed of light, and $\theta$ the viewing
  angle. The exact value of the $\delta$ exponent depends on the geometry of
  the emitting region.}. 

The RLF of RL-AGNs extending to the lowest value of radio luminosity has been
obtained by \citet{mau07}, covering the range $\log P_{\rm r} = 20.4 - 26.4 \,
\rm [W \, Hz^{-1}]$. At low luminosities, the RLF does not significantly depart
from a power law, and their results suggest that any break must be located
below $\log P_{r} \sim 21.6\, \rm W \, Hz^{-1}$, where the limited number of
observed objects makes the RLF shape uncertain.  On the other hand, the number
density of radio emitting AGNs cannot exceed that of their potential hosts:
this argument has been used by \citet{capetti2015}, requiring a minimum black
hole mass of $\log \,(M_{\rm BH}/M_\odot) > 7.5$ for the host of RL-AGNs. With
this approach they found that the break in the RLF of RL-AGNs must be located
at $P_{\rm r} \ga 10^{20.5} \, \rm W \, Hz^{-1}$.

The ratio between the break powers in the RLFs of BL~Lacs and RL-AGNs is a
factor $\sim 10^3 - 10^4$.  If adopting $p=3$, this suggests a value of the
Doppler factor $\delta_{\rm max} \sim 10 - 20$, consistent with the typical
results from the observations and the model predictions
\citep[e.g.,][]{ghi93,sav10}.

\section{Summary and conclusions}

In this paper we presented a new strategy for recognizing BL Lac objects, based
on the combination of optical indices derived from the SDSS spectra with MIR
colors from the AllWISE survey. This enabled us to separate BL Lac objects
from low-luminosity AGNs and star-forming galaxies. Our method is particularly
useful for identifying LPBL, where the traditional optical non-thermal signature
is reduced thanks to the dominant host-galaxy emission. We applied it to the
sample of 18~286 radio sources selected by BH12 by combing the SDSS DR 7
spectra with the NVSS and FIRST surveys.  We showed that in the Dn(4000)
versus W2$-$W3 plane, BL Lacs fall in a region (defined as ``forbidden zone'')
that is scarcely populated by the general galaxies population. The main contaminants
are the A+E (post-starburst) galaxies that can be easily removed with the aid
of their characteristic pronounced Balmer absorption lines. When considering
low luminosities, the use of the Dn(4000) index alone is not sufficient to
obtain a clean sample. In fact, the constraint based on the W2$-$W3 color
allowed us to remove 54 objects (most of which are low-luminosity LINERs
and/or Seyfert galaxies) that would instead be selected as BL Lac
candidates. Including the MIR data thus substantially improves the
quality of the resulting BL Lac sample.

We isolated 36 BL Lac candidates up to $z=0.15$. All but one are associated
with luminous elliptical galaxies with radio powers ranging from $\log L_{\rm
  r} =39.2$ to 41.5 [$\ergs$]. Fifteen of the selected objects are new
identifications, most of which have an X-ray counterpart.

We discussed various issues concerning our sample completeness.  The radio
flux threshold does not have a significant impact on our results.  As for the
optical selection, most of the expected RL-AGN hosts are included in the
MPA-JHU catalog. The only exceptions are the low-luminosity tail close to the
upper redshift boundary and the most luminous galaxies located at very low
redshift.  We then set a more stringent lower limit of $z=0.05$ on the
galaxies considered.  We used the BZCAT for an external check of our
results. We recovered all the known objects in the catalog up to $z=0.1$,
while we lost less than 10\% of them in the $0.1 < z <0.15$ range. In
particular, only two very bright BL Lacs are lost because of their point-like
appearance, which caused these objects to be misclassified as ``STAR" in the
SDSS. We also tested the effects of the criterion based on the W2$-$W3
  threshold. Out of the 54 sources mentioned above, we find five objects that
  conform to the general properties of BL~Lacs (since hosted in
  elliptical galaxies and having low lines EW). However, they are of
  uncertain nature because they lie in a region of the W2$-$W3 vs. Dn(4000)
  plane populated by other classes of AGNs.

We simulated the selection function of BL Lacs within our sample. The key
  ingredient is the assumption on the distribution of the radio-to-optical
  spectral index for which we adopted the values derived from our candidates.
  Since they agree with those found in the literature, our results
  are not significantly affected by selections biases. The sample completeness
  reaches $\sim 50\%$ at $\log L_{\rm r} \sim 39.7$ [$\ergs$].

The selected BL Lac candidates were used to explore the BL Lacs RLF at low
radio powers.  Since we are dealing with an essentially volume-limited sample at
low redshift, we can neglect evolutionary effects on the RLF.  This can be obtained 
simply by counting all objects at the various radio luminosities within
the considered volume. We found good agreement with the results
  by previous authors for $\log P_{\rm r} > 24.5 \, \rm [W \, Hz^{-1}]$.

However, we found a dramatic paucity of LPBL with respect to the
extrapolation of the BL~Lacs RLF, a result that is altered neither by the
  inclusion of the five potential BL~Lacs nor by the uncertainties on the
  selection function. This occurs well above the power at which our selection
function falls significantly and requires a break in the RLF.  We estimated
that the break is located at $\log P_{\rm r} \sim 24.5 \, \rm [W \,
  Hz^{-1}]$. The slope of the RLF below the break is consistent with a zero
value, implying the presence of a peak in the number density distribution of
BL Lacs.

In the framework of the AGN unification scheme, BL Lacs are the beamed
counterparts of low-power RL-AGNs, so that a link between their RLFs is
expected.  In particular, a low-power break in the parent population
translates into a change in slope in the RLF of BL Lacs.  Indeed, there is
evidence of a low-luminosity break in the RL-AGN RLF located about three to four
  orders of magnitude below what we found for BL Lacs. This
  corresponds to a maximum Doppler factor of $\sim 10$ -- 20, in agreement
with the observations and model predictions.

In the near future, we plan to do a detailed analysis of the whole BL Lac candidate sample derived in this paper with two aims: i) to confirm their nature as genuine BL Lacs and ii) to characterize their broad-band properties.
In particular, we need to obtain high-resolution radio data in order to isolate their nuclear emission and X-ray data to explore the high-energy non-thermal emission.

\begin{acknowledgements}
We are deeply indebted to the anonymous referee for the careful reading
of the manuscript and for the many useful and insightful suggestions.

This research made use of the NASA/ IPAC Infrared Science Archive and
Extragalactic Database (NED), which are operated by the Jet Propulsion
Laboratory, California Institute of Technology, under contract with the
National Aeronautics and Space Administration.

Funding for SDSS-III has been provided by the Alfred P. Sloan Foundation, the
Participating Institutions, the National Science Foundation, and the
U.S. Department of Energy Office of Science. The SDSS-III web site is
http://www.sdss3.org/.  SDSS-III is managed by the Astrophysical Research
Consortium for the Participating Institutions of the SDSS-III Collaboration,
including the University of Arizona, the Brazilian Participation Group,
Brookhaven National Laboratory, University of Cambridge, Carnegie Mellon
University, University of Florida, the French Participation Group, the German
Participation Group, Harvard University, the Instituto de Astrofisica de
Canarias, the Michigan State/Notre Dame/JINA Participation Group, Johns
Hopkins University, Lawrence Berkeley National Laboratory, Max Planck
Institute for Astrophysics, Max Planck Institute for Extraterrestrial Physics,
New Mexico State University, New York University, Ohio State University,
Pennsylvania State University, University of Portsmouth, Princeton University,
the Spanish Participation Group, University of Tokyo, University of Utah,
Vanderbilt University, University of Virginia, University of Washington, and
Yale University.

This publication makes use of data products from the Two Micron All Sky Survey, which is a joint project of the University of Massachusetts and the Infrared Processing and Analysis Center/California Institute of Technology, funded by the National Aeronautics and Space Administration and the National Science Foundation.

Part of this work is based on archival data, software or online services provided by the ASI Science Data Center (ASDC). 

\end{acknowledgements}

\bibliographystyle{aa}

\begin{thebibliography}{48}
\expandafter\ifx\csname natexlab\endcsname\relax\def\natexlab#1{#1}\fi

\bibitem[{{Abdo} {et~al.}(2010){Abdo}, {Ackermann}, {Ajello}, {Atwood},
  {Axelsson}, {Baldini}, {Ballet}, {Barbiellini}, {Bastieri}, {Bechtol},
  {Bellazzini}, {Berenji}, {Bloom}, {Bonamente}, {Borgland}, {Bregeon}, {Brez},
  {Brigida}, {Bruel}, \& {Burnett}}]{abdo2010}
{Abdo}, A.~A., {Ackermann}, M., {Ajello}, M., {et~al.} 2010, \apj, 708, 1310

\bibitem[{{Aller} {et~al.}(1996){Aller}, {Aller}, \& {Hughes}}]{all96}
{Aller}, M.~F., {Aller}, H.~D., \& {Hughes}, P.~A. 1996, in Astronomical
  Society of the Pacific Conference Series, Vol. 110, Blazar Continuum
  Variability, ed. H.~R. {Miller}, J.~R. {Webb}, \& J.~C. {Noble}, 193

\bibitem[{{Antonucci} \& {Ulvestad}(1985)}]{ant85}
{Antonucci}, R.~R.~J. \& {Ulvestad}, J.~S. 1985, \apj, 294, 158

\bibitem[{{Baldi} \& {Capetti}(2010)}]{baldi10b}
{Baldi}, R.~D. \& {Capetti}, A. 2010, \aap, 519, A48+

\bibitem[{{Baldi} {et~al.}(2015){Baldi}, {Capetti}, \& {Giovannini}}]{baldi15}
{Baldi}, R.~D., {Capetti}, A., \& {Giovannini}, G. 2015, \aap, 576, A38

\bibitem[{{Balogh} {et~al.}(1999){Balogh}, {Morris}, {Yee}, {Carlberg}, \&
  {Ellingson}}]{bal99}
{Balogh}, M.~L., {Morris}, S.~L., {Yee}, H.~K.~C., {Carlberg}, R.~G., \&
  {Ellingson}, E. 1999, \apj, 527, 54

\bibitem[{{Best} \& {Heckman}(2012)}]{bes12alias}
{Best}, P.~N. \& {Heckman}, T.~M. 2012, \mnras, 421, 1569 (BH12)

\bibitem[{{Brinchmann} {et~al.}(2004){Brinchmann}, {Charlot}, {White},
  {Tremonti}, {Kauffmann}, {Heckman}, \& {Brinkmann}}]{bri04}
{Brinchmann}, J., {Charlot}, S., {White}, S.~D.~M., {et~al.} 2004, \mnras, 351,
  1151

\bibitem[{{Capetti} \& {Raiteri}(2015)}]{capetti2015}
{Capetti}, A. \& {Raiteri}, C.~M. 2015, \mnras

\bibitem[{{Donato} {et~al.}(2001){Donato}, {Ghisellini}, {Tagliaferri}, \&
  {Fossati}}]{donato01}
{Donato}, D., {Ghisellini}, G., {Tagliaferri}, G., \& {Fossati}, G. 2001, \aap,
  375, 739

\bibitem[{{Fossati} {et~al.}(1998){Fossati}, {Maraschi}, {Celotti}, {Comastri},
  \& {Ghisellini}}]{fos98}
{Fossati}, G., {Maraschi}, L., {Celotti}, A., {Comastri}, A., \& {Ghisellini},
  G. 1998, \mnras, 299, 433

\bibitem[{{Ghisellini} {et~al.}(1993){Ghisellini}, {Padovani}, {Celotti}, \&
  {Maraschi}}]{ghi93}
{Ghisellini}, G., {Padovani}, P., {Celotti}, A., \& {Maraschi}, L. 1993, \apj,
  407, 65

\bibitem[{{Giommi} {et~al.}(2012{\natexlab{a}}){Giommi}, {Padovani}, {Polenta},
  {Turriziani}, {D'Elia}, \& {Piranomonte}}]{gio12a}
{Giommi}, P., {Padovani}, P., {Polenta}, G., {et~al.} 2012{\natexlab{a}},
  \mnras, 420, 2899

\bibitem[{{Giommi} {et~al.}(2012{\natexlab{b}}){Giommi}, {Polenta},
  {L{\"a}hteenm{\"a}ki}, {Thompson}, {Capalbi}, {Cutini}, {Gasparrini},
  {Gonz{\'a}lez-Nuevo}, {Le{\'o}n-Tavares}, {L{\'o}pez-Caniego}, {Mazziotta},
  {Monte}, {Perri}, {Rain{\`o}}, {Tosti}, {Tramacere}, {Verrecchia}, {Aller},
  {Aller}, {Angelakis}, {Bastieri}, {Berdyugin}, {Bonaldi}, {Bonavera},
  {Burigana}, {Burrows}, {Buson}, {Cavazzuti}, {Chincarini}, {Colafrancesco},
  {Costamante}, {Cuttaia}, {D'Ammando}, {de Zotti}, {Frailis}, {Fuhrmann},
  {Galeotta}, {Gargano}, {Gehrels}, {Giglietto}, {Giordano}, {Giroletti},
  {Keih{\"a}nen}, {King}, {Krichbaum}, {Lasenby}, {Lavonen}, {Lawrence},
  {Leto}, {Lindfors}, {Mandolesi}, {Massardi}, {Max-Moerbeck}, {Michelson},
  {Mingaliev}, {Natoli}, {Nestoras}, {Nieppola}, {Nilsson}, {Partridge},
  {Pavlidou}, {Pearson}, {Procopio}, {Rachen}, {Readhead}, {Reeves}, {Reimer},
  {Reinthal}, {Ricciardi}, {Richards}, {Riquelme}, {Saarinen}, {Sajina},
  {Sandri}, {Savolainen}, {Sievers}, {Sillanp{\"a}{\"a}}, {Sotnikova},
  {Stevenson}, {Tagliaferri}, {Takalo}, {Tammi}, {Tavagnacco}, {Terenzi},
  {Toffolatti}, {Tornikoski}, {Trigilio}, {Turunen}, {Umana}, {Ungerechts},
  {Villa}, {Wu}, {Zacchei}, {Zensus}, \& {Zhou}}]{giommi12}
{Giommi}, P., {Polenta}, G., {L{\"a}hteenm{\"a}ki}, A., {et~al.}
  2012{\natexlab{b}}, \aap, 541, A160

\bibitem[{{Goto} {et~al.}(2003){Goto}, {Nichol}, {Okamura}, {Sekiguchi},
  {Miller}, {Bernardi}, {Hopkins}, {Tremonti}, {Connolly}, {Castander},
  {Brinkmann}, {Fukugita}, {Harvanek}, {Ivezic}, {Kleinman}, {Krzesinski},
  {Long}, {Loveday}, {Neilsen}, {Newman}, {Nitta}, {Snedden}, \&
  {Subbarao}}]{got03}
{Goto}, T., {Nichol}, R.~C., {Okamura}, S., {et~al.} 2003, \pasj, 55, 771

\bibitem[{{Kellermann} {et~al.}(2004){Kellermann}, {Lister}, {Homan},
  {Vermeulen}, {Cohen}, {Ros}, {Kadler}, {Zensus}, \& {Kovalev}}]{kel04}
{Kellermann}, K.~I., {Lister}, M.~L., {Homan}, D.~C., {et~al.} 2004, \apj, 609,
  539

\bibitem[{{Konigl}(1981)}]{kon81}
{Konigl}, A. 1981, \apj, 243, 700

\bibitem[{{Landt} {et~al.}(2002){Landt}, {Padovani}, \& {Giommi}}]{landt02}
{Landt}, H., {Padovani}, P., \& {Giommi}, P. 2002, \mnras, 336, 945

\bibitem[{{Liuzzo} {et~al.}(2013){Liuzzo}, {Giroletti}, {Giovannini},
  {Boccardi}, {Tamburri}, {Taylor}, {Casadio}, {Kadler}, {Tosti}, \&
  {Mignano}}]{liuzzo2013}
{Liuzzo}, E., {Giroletti}, M., {Giovannini}, G., {et~al.} 2013, \aap, 560, A23

\bibitem[{{March{\~a}} \& {Caccianiga}(2013)}]{marcha2013}
{March{\~a}}, M.~J.~M. \& {Caccianiga}, A. 2013, \mnras, 430, 2464

\bibitem[{{Marcha} {et~al.}(1996){Marcha}, {Browne}, {Impey}, \&
  {Smith}}]{marcha1996}
{Marcha}, M.~J.~M., {Browne}, I.~W.~A., {Impey}, C.~D., \& {Smith}, P.~S. 1996,
  \mnras, 281, 425

\bibitem[{{Massaro} {et~al.}(2009){Massaro}, {Giommi}, {Leto}, {Marchegiani},
  {Maselli}, {Perri}, {Piranomonte}, \& {Sclavi}}]{mas09}
{Massaro}, E., {Giommi}, P., {Leto}, C., {et~al.} 2009, \aap, 495, 691

\bibitem[{{Massaro} {et~al.}(2011){Massaro}, {D'Abrusco}, {Ajello}, {Grindlay},
  \& {Smith}}]{masf11}
{Massaro}, F., {D'Abrusco}, R., {Ajello}, M., {Grindlay}, J.~E., \& {Smith},
  H.~A. 2011, \apjl, 740, L48

\bibitem[{{Mauch} \& {Sadler}(2007)}]{mau07}
{Mauch}, T. \& {Sadler}, E.~M. 2007, \mnras, 375, 931

\bibitem[{{Montero-Dorta} \& {Prada}(2009)}]{mon09}
{Montero-Dorta}, A.~D. \& {Prada}, F. 2009, \mnras, 399, 1106

\bibitem[{{Padovani} {et~al.}(2007){Padovani}, {Giommi}, {Landt}, \&
  {Perlman}}]{pad07}
{Padovani}, P., {Giommi}, P., {Landt}, H., \& {Perlman}, E.~S. 2007, \apj, 662,
  182

\bibitem[{{Padovani} {et~al.}(2003){Padovani}, {Perlman}, {Landt}, {Giommi}, \&
  {Perri}}]{pad03}
{Padovani}, P., {Perlman}, E.~S., {Landt}, H., {Giommi}, P., \& {Perri}, M.
  2003, \apj, 588, 128

\bibitem[{{Plotkin} {et~al.}(2010){Plotkin}, {Anderson}, {Brandt},
  {Diamond-Stanic}, {Fan}, {Hall}, {Kimball}, {Richmond}, {Schneider},
  {Shemmer}, {Voges}, {York}, {Bahcall}, {Snedden}, {Bizyaev}, {Brewington},
  {Malanushenko}, {Malanushenko}, {Oravetz}, {Pan}, \& {Simmons}}]{plo10}
{Plotkin}, R.~M., {Anderson}, S.~F., {Brandt}, W.~N., {et~al.} 2010, \aj, 139,
  390

\bibitem[{{Polletta} {et~al.}(2007){Polletta}, {Tajer}, {Maraschi},
  {Trinchieri}, {Lonsdale}, {Chiappetti}, {Andreon}, {Pierre}, {Le F{\`e}vre},
  \& {Zamorani}}]{pol07}
{Polletta}, M., {Tajer}, M., {Maraschi}, L., {et~al.} 2007, \apj, 663, 81

\bibitem[{{Raiteri} {et~al.}(2014){Raiteri}, {Villata}, {Carnerero},
  {Acosta-Pulido}, {Larionov}, {D'Ammando}, {Ar{\'e}valo}, {Arkharov}, {Bueno
  Bueno}, {Di Paola}, {Efimova}, {Gonz{\'a}lez-Morales}, {Gorshanov},
  {Grinon-Marin}, {L{\'a}zaro}, {Manilla-Robles}, {Pastor Yabar}, {Puerto
  Gim{\'e}nez}, \& {Velasco}}]{rai14}
{Raiteri}, C.~M., {Villata}, M., {Carnerero}, M.~I., {et~al.} 2014, \mnras,
  442, 629

\bibitem[{{Sadler} {et~al.}(2014){Sadler}, {Ekers}, {Mahony}, {Mauch}, \&
  {Murphy}}]{sadler14}
{Sadler}, E.~M., {Ekers}, R.~D., {Mahony}, E.~K., {Mauch}, T., \& {Murphy}, T.
  2014, \mnras, 438, 796

\bibitem[{{Sambruna} {et~al.}(1996){Sambruna}, {Maraschi}, \&
  {Urry}}]{sambruna96}
{Sambruna}, R.~M., {Maraschi}, L., \& {Urry}, C.~M. 1996, \apj, 463, 444

\bibitem[{{Savolainen} {et~al.}(2010){Savolainen}, {Homan}, {Hovatta},
  {Kadler}, {Kovalev}, {Lister}, {Ros}, \& {Zensus}}]{sav10}
{Savolainen}, T., {Homan}, D.~C., {Hovatta}, T., {et~al.} 2010, \aap, 512, A24

\bibitem[{{Smith}(1996)}]{smi96}
{Smith}, A.~G. 1996, in Astronomical Society of the Pacific Conference Series,
  Vol. 110, Blazar Continuum Variability, ed. H.~R. {Miller}, J.~R. {Webb}, \&
  J.~C. {Noble}, 3

\bibitem[{{Stickel} {et~al.}(1991){Stickel}, {Fried}, {Kuehr}, {Padovani}, \&
  {Urry}}]{sti91}
{Stickel}, M., {Fried}, J.~W., {Kuehr}, H., {Padovani}, P., \& {Urry}, C.~M.
  1991, \apj, 374, 431

\bibitem[{{Stocke} {et~al.}(1991){Stocke}, {Morris}, {Gioia}, {Maccacaro},
  {Schild}, {Wolter}, {Fleming}, \& {Henry}}]{sto91}
{Stocke}, J.~T., {Morris}, S.~L., {Gioia}, I.~M., {et~al.} 1991, \apjs, 76, 813

\bibitem[{{Strauss} {et~al.}(2002){Strauss}, {Weinberg}, {Lupton}, {Narayanan},
  {Annis}, {Bernardi}, {Blanton}, {Burles}, {Connolly}, {Dalcanton}, {Doi},
  {Eisenstein}, {Frieman}, {Fukugita}, {Gunn}, {Ivezi{\'c}}, {Kent}, {Kim},
  {Knapp}, {Kron}, {Munn}, {Newberg}, {Nichol}, {Okamura}, {Quinn}, {Richmond},
  {Schlegel}, {Shimasaku}, {SubbaRao}, {Szalay}, {Vanden Berk}, {Vogeley},
  {Yanny}, {Yasuda}, {York}, \& {Zehavi}}]{str02}
{Strauss}, M.~A., {Weinberg}, D.~H., {Lupton}, R.~H., {et~al.} 2002, \aj, 124,
  1810

\bibitem[{{Tremaine} {et~al.}(2002){Tremaine}, {Gebhardt}, {Bender}, {Bower},
  {Dressler}, {Faber}, {Filippenko}, {Green}, {Grillmair}, {Ho}, {Kormendy},
  {Lauer}, {Magorrian}, {Pinkney}, \& {Richstone}}]{tre02}
{Tremaine}, S., {Gebhardt}, K., {Bender}, R., {et~al.} 2002, \apj, 574, 740

\bibitem[{{Tremonti} {et~al.}(2004){Tremonti}, {Heckman}, {Kauffmann},
  {Brinchmann}, {Charlot}, {White}, {Seibert}, {Peng}, {Schlegel}, {Uomoto},
  {Fukugita}, \& {Brinkmann}}]{tre04}
{Tremonti}, C.~A., {Heckman}, T.~M., {Kauffmann}, G., {et~al.} 2004, \apj, 613,
  898

\bibitem[{{Urry} \& {Padovani}(1991)}]{urr91}
{Urry}, C.~M. \& {Padovani}, P. 1991, \apj, 371, 60

\bibitem[{{Urry} \& {Padovani}(1995)}]{urr95}
{Urry}, C.~M. \& {Padovani}, P. 1995, \pasp, 107, 803

\bibitem[{{Urry} {et~al.}(1991){Urry}, {Padovani}, \& {Stickel}}]{urr91a}
{Urry}, C.~M., {Padovani}, P., \& {Stickel}, M. 1991, \apj, 382, 501

\bibitem[{{Urry} \& {Shafer}(1984)}]{urr84}
{Urry}, C.~M. \& {Shafer}, R.~A. 1984, \apj, 280, 569

\bibitem[{{Wagner} {et~al.}(1996){Wagner}, {Witzel}, {Heidt}, {Krichbaum},
  {Qian}, {Quirrenbach}, {Wegner}, {Aller}, {Aller}, {Anton}, {Appenzeller},
  {Eckart}, {Kraus}, {Naundorf}, {Kneer}, {Steffen}, \& {Zensusj}}]{wag96}
{Wagner}, S.~J., {Witzel}, A., {Heidt}, J., {et~al.} 1996, \aj, 111, 2187

\bibitem[{{Wolter} {et~al.}(1994){Wolter}, {Caccianiga}, {della Ceca}, \&
  {Maccacaro}}]{wol94}
{Wolter}, A., {Caccianiga}, A., {della Ceca}, R., \& {Maccacaro}, T. 1994,
  \apj, 433, 29

\bibitem[{{Worthey} \& {Ottaviani}(1997)}]{worthey97}
{Worthey}, G. \& {Ottaviani}, D.~L. 1997, \apjs, 111, 377

\bibitem[{{Wright} {et~al.}(2010){Wright}, {Eisenhardt}, {Mainzer}, {Ressler},
  {Cutri}, {Jarrett}, {Kirkpatrick}, {Padgett}, {McMillan}, \&
  {Skrutskie}}]{wri10}
{Wright}, E.~L., {Eisenhardt}, P.~R.~M., {Mainzer}, A.~K., {et~al.} 2010, \aj,
  140, 1868

\bibitem[{{Zehavi} {et~al.}(2002){Zehavi}, {Blanton}, {Frieman}, {Weinberg},
  {Mo}, {Strauss}, {Anderson}, {Annis}, {Bahcall}, {Bernardi}, {Briggs},
  {Brinkmann}, {Burles}, {Carey}, {Castander}, {Connolly}, {Csabai},
  {Dalcanton}, {Dodelson}, {Doi}, {Eisenstein}, {Evans}, {Finkbeiner},
  {Friedman}, {Fukugita}, {Gunn}, {Hennessy}, {Hindsley}, {Ivezi{\'c}}, {Kent},
  {Knapp}, {Kron}, {Kunszt}, {Lamb}, {Leger}, {Long}, {Loveday}, {Lupton},
  {McKay}, {Meiksin}, {Merrelli}, {Munn}, {Narayanan}, {Newcomb}, {Nichol},
  {Owen}, {Peoples}, {Pope}, {Rockosi}, {Schlegel}, {Schneider}, {Scoccimarro},
  {Sheth}, {Siegmund}, {Smee}, {Snir}, {Stebbins}, {Stoughton}, {SubbaRao},
  {Szalay}, {Szapudi}, {Tegmark}, {Tucker}, {Uomoto}, {Vanden Berk}, {Vogeley},
  {Waddell}, {Yanny}, \& {York}}]{zeh02}
{Zehavi}, I., {Blanton}, M.~R., {Frieman}, J.~A., {et~al.} 2002, \apj, 571, 172

\end{thebibliography}

\appendix

\section{Details of the comparison with BZCAT.}
\label{abzcat}
In the BZCAT list there are 34 BZB with $0.05 < z \le 0.1$,
nine of which are located in the area covered by both FIRST and the spectroscopic DR 7.
Seven of them are in our list of BL Lacs candidates. The remaining two are:

\noindent
- BZBJ1216+0929, which  was selected as spectroscopic target, but its spectrum was not obtained owing to a fiber collision; and

\noindent
- BZBJ1426+2415, which  shows a featureless spectrum; the lack of any signature from the galaxy emission cast doubts on
the accuracy of its tentative redshift estimate, $z=0.055$; indeed, the DR 7 value is $z=2.243$.

A check on the BZU is more problematic, because most of them are FSRQs.
There are 24 BZU in BZCAT at $0.05 < z \le 0.1$.
Nine of them are in our survey area, but two lack spectra because of fiber collision.
The other seven have strong emission lines.

In the redshift range $0.1 < z \le 0.15$, we found 59 BZB in BZCAT, 26 of which
are located in the area covered by both the spectroscopic DR7 and FIRST.  When
excluding three objects that have no spectrum because of fiber collision and one
object whose spectrum has a defect in the H$\alpha$ region, we are left with
22 sources, 15 of which belong to the BH12 sample.  The other seven are lost
because:

\noindent
- one object has no spectrum because it is too faint, but we know from BOSS that it is a spiral galaxy;

\noindent
- three are featureless objects whose SDSS redshift is higher than 0.15;

\noindent
- one is a strong-lined object reaching an EW of $\sim 10$ \AA;

\noindent
- two (BZBJ0809+5218 and BZBJ1221+2813-W Comae) are classified as ``STARS" in
the SDSS and thus do not enter in the main galaxy sample analyzed by BH12;
they have featureless spectra, and $\log L_{\rm r}=41.11$ and 41.54,
respectively.

Among the 15 sources included in the BH12 sample, 13 are in our list of BL
Lacs candidates, while two sources (BZBJ0810+4911 and BZBJ1404+2701) were not
selected by our method because of a high Dn(4000).  They have $\log L_{\rm
  r}=39.50$ and 40.39, respectively. The first one was identified as a BL Lac
because of $\gamma$-ray emission detected by the EGRET instrument on-board the
Compton Gamma-Ray Observatory, but it is not present in the third Fermi Large
Area Telescope source catalog. Furthermore, \citet{liuzzo2013} did not detect
it with the VLA at 8.4 GHz, so we consider it a dubious association.

The above analysis leads to the conclusion that in the $0.1 < z \le 0.15$
redshift range, our selection method recovered all BZB but two sources
characterized by a high Dn(4000) value. Another two objects were lost because they
are so bright that they are not classified as galaxies and so do not enter the
BH12 sample. Another four objects lack complete spectral information.

Among the BZU, there are 23 objects in BZCAT at $0.1 < z \le 0.15$, 14 that 
are in the region covered by both the spectroscopic DR7 and FIRST.  One
object has no spectrum because of fiber collision, two have strong emission
lines, 11 are in the BH12 sample, but only three have rest-frame EWs less than 5
\AA\ and only one, BZUJ0936+0509, is selected by our method (Table
\ref{bzall}).  The other two are BZUJ0820+4853 and BZUJ0948+5535.  The former
was discarded because of a high Dn(4000) of 2.01, while the latter lies
  in the ``wedge" region of Fig. 7. They have $\log L_{\rm r}=40.65$ and 40.12,
respectively.

The analysis of the BZU in the $0.1 < z \le 0.15$ redshift range thus reveals that our methods did not recover two sources out of 14. One other object was missed because of fiber collision.

\end{document}